\documentclass[twoside]{article}
\usepackage{qic,epsfig}
\usepackage[utf8]{inputenc}
\usepackage{fullpage}
\usepackage{setspace}

\usepackage{cite}
\usepackage{graphicx}
\usepackage{listings}
\usepackage{tabularx}
\usepackage{xcolor}
\usepackage{tikz}
\usepackage{courier}
\usepackage{url}
\usepackage{verbatim}
\usepackage[caption=false]{subfig}

\begin{document}

\usetikzlibrary{arrows,shapes,trees,backgrounds, shapes}
\tikzset{cross/.style={cross out, draw, 
         minimum size=2*(#1-\pgflinewidth), 
         inner sep=0pt, outer sep=0pt},
cross/.default={4pt}}

\centerline{\bf
\uppercase{Local Decoders for the 2D and 4D Toric Code}}

\vspace*{0.37truein}
\centerline{\footnotesize
\textsc{Nikolas P. Breuckmann}}
\vspace*{0.015truein}
\centerline{\footnotesize\it JARA Institute for Quantum Information, RWTH Aachen, Otto-Blumenthal-Stra{\ss}e 20}
\baselineskip=10pt
\centerline{\footnotesize\it Aachen, Nordrhein-Westfalen, 52074 Germany}
\vspace*{10pt}
\centerline{\footnotesize 
\textsc{Kasper Duivenvoorden}}
\vspace*{0.015truein}
\centerline{\footnotesize\it JARA Institute for Quantum Information, RWTH Aachen, Otto-Blumenthal-Stra{\ss}e 20}
\baselineskip=10pt
\centerline{\footnotesize\it Aachen, Nordrhein-Westfalen, 52074 Germany}
\vspace*{10pt}
\centerline{\footnotesize 
\textsc{Dominik Michels}}
\vspace*{0.015truein}
\centerline{\footnotesize\it JARA Institute for Quantum Information, RWTH Aachen, Otto-Blumenthal-Stra{\ss}e 20}
\baselineskip=10pt
\centerline{\footnotesize\it Aachen, Nordrhein-Westfalen, 52074 Germany}
\vspace*{10pt}
\centerline{\footnotesize 
\textsc{Barbara M. Terhal}}
\vspace*{0.015truein}
\centerline{\footnotesize\it JARA Institute for Quantum Information, RWTH Aachen, Otto-Blumenthal-Stra{\ss}e 20}
\baselineskip=10pt
\centerline{\footnotesize\it Aachen, Nordrhein-Westfalen, 52074 Germany}
\vspace*{0.225truein}
\publisher{(received date)}{(revised date)}

\vspace*{0.21truein}

\begin{abstract}
 We analyze the performance of decoders for the 2D and 4D toric code which are local by construction. The 2D decoder is a cellular automaton decoder formulated by Harrington \cite{thesis:harrington} which explicitly has a finite speed of communication and computation. For a model of independent $X$ and $Z$ errors and faulty syndrome measurements with identical probability we report a threshold of $0.133\%$ for this Harrington decoder. We implement a decoder for the 4D toric code which is based on a decoder by Hastings \cite{hastings}. Incorporating a method for handling faulty syndromes we estimate a threshold of $1.59\%$ for the same noise model as in the 2D case. We compare the performance of this decoder with a decoder based on a 4D version of Toom's cellular automaton rule as well as the decoding method suggested by Dennis {\em et al.} \cite{DKLP}.
\end{abstract}

\tableofcontents

\onehalfspacing

\section{Introduction}
In the last 10 years the toric or surface code has become one of the most viable candidates for storing quantum information reliably in a two-dimensional system \cite{DKLP}. The toric code combines a high noise threshold, almost $1\%$ for circuit-based noise \cite{fowler2009high}, with an overhead-efficient way of performing Clifford gates such as Hadamard and CNOT, as well as a methodology for implementing non-Clifford gates via magic-state-distillation, see e.g. the review in \cite{FMMC:review}. 

The noise threshold of any quantum error correction code depends on the way in which errors are inferred from information gained about the errors; this information is the so-called error {\em syndrome} obtained via parity checks measurements. This classical inference problem is the computational problem solved by the classical decoder.

It is known that for storing information in quantum memory or performing Clifford gates the classical decoder can operate {\em off-line}, meaning that it can process the syndrome information after the entire computation has finished, see \cite{FMMC:review,terhal:RMP}. This implies that demands on decoding speed are relaxed and surface code decoding can take place via the non-local minimum weight matching method, producing close-to-optimal noise thresholds. However, on-line decoding will be essential for quantum computation when applying non-Clifford gates because one has to know the logical Pauli frame in time in order not to produce logical Clifford errors, see e.g. the discussion in \cite{terhal:RMP}. Practically speaking, the I/O demands and speed of the decoder are expected to be high. For example, a QEC clock-cycle for superconducting hardware can be estimated as 5 MHz, which for a qubit encoded in a $L \times L$ toric code lattice with $L=100$, leads to a data rate of 100Gbit/sec from the superconducting chip into a classical decoder implemented in software or dedicated hardware. 

At a fundamental level, universal quantum computation via the 2D surface code necessitates at least a parallel-operating decoder whose rate in processing syndrome information is independent of code size. A parallel implementation of the minimum weight matching algorithm for decoding has been considered in \cite{fowler:O(1)}, claiming O(1) processing time on average. In addition, renormalization group (RG) decoders exist, e.g. \cite{DP:3Ddecoding}, whose operation is naturally parallel over the lattice, which can be shown to run in $O(\log L)$ parallel time for a $L \times L$ lattice.
Ideally, the decoding task is executed by an 2D array of spatially \textit{local} computing cells which communicate and compute at finite speed.  By varying computation and/or communication rates (as compared to the syndrome acquisition rate) such decoder can be more or less local.

The question of the existence of such local decoder for the 2D toric code is closely related to the question of the existence of a local decoder for the 1D classical repetition code: both models require the decoder to have a hierarchical structure in which the removal of errors on length scale $l$ requires communication at length scale $l$ and thus time scaling with $l$. The 1D problem was solved by Gacs \cite{GACS198615} by formulating a hierarchically-operating noisy cellular automaton decoder. In his PhD thesis Jim Harrington describes a 2D cellular automaton decoder for the toric code in analogy
with the Gacs work. He provides arguments that the local decoder has a noise threshold (equal to at least $2.4\times10^{-11}$). He also provides numerical evidence concerning the actual memory time of the stored information suggesting a substantially larger threshold of $O(10^{-3})-O(10^{-4})$.
 
In Sections \ref{sec:harrington} and \ref{sec:numerics} of this paper we re-examine the Harrington decoder and numerically estimate the best achievable memory (life) times and noise threshold.   We give a self-contained explanation of the decoder and discuss the effect of various parameters which define the decoder. We vary the speed of the classical cellular automaton as compared to the syndrome acquisition rate and study the effect on the memory time. We give evidence supporting a threshold of $0.133\%$ when various parameters are optimized.  

In Section \ref{sec:physics} we discuss why the principle underlying Harrington's decoder may allow one to make the memory size of each cell $O(1)$ but at the cost of high complexity in formulating the cellular automaton rules. In this Section we also discuss the question of passive 2D quantum error correction from the perspective of concatenated coding. \\

\subsection{4D Toric Code}
The disadvantages of the 2D toric code, namely the need for non-local decoding in space-time, are known to be absent for the 4D toric code \cite{DKLP} which can be decoded via {\em single-shot means}. In practice this means that reliable ancillary (qu)bits can be used to obtain error information, compute and implement physical error correction locally (that is, w.r.t. the 4D hypercubic lattice), without the requirement for a hierarchical structure which corrects errors at higher length scales. Even though the 4D toric code may be an impractical choice as the bottom code in a coding hardware when only 1D or 2D local connectivity is available, it could still have uses as a top code in which the logical qubits encoded by a 2D surface code are used as elementary qubits in this 4D code. Whether the 4D toric code (or even a 4D hyperbolic code family \cite{guth_lubotzky,hastings} with a constant encoding rate $R=\frac{k}{n}$) may have any practical use, depends on its noise threshold and its decoding method. In Section \ref{sec:4Ddecoding} and Section \ref{sec:numerics4D} we report our results on implementing various local single-shot decoders for the 4D toric code. This includes a decoder based on the work in \cite{hastings} for the 4D toric code, adapted to be resilient to measurement errors as well as cellular automaton decoders first considered in \cite{DKLP}.

In the next section we start by reviewing the definition of the toric codes and some of the previous work on decoders and noise thresholds. We end the paper with a discussion and possible future work.

\section{The Toric Codes}\label{sec:toric_codes}
The 2D (resp. 4D) toric code is defined on a 2-dimensional square (resp. 4-dimensional hypercubic) lattice with linear size $L$ and periodic boundary conditions in all directions. On the 2D lattice one can define vertices, edges and faces and qubits are defined on the $n=2L^2$ edges (2 per unit cell).
On the 4D lattice one can define vertices, edges, faces and cubes and qubits are defined on the $6L^4$ faces (6 per unit cell). The generators of the 2D toric code stabilizer group are $X$ and $Z$-parity check operators. Each parity $Z$-check corresponds to the edge boundary of a face, it acts as $Z$ on the four qubits surrounding a face. Each parity $X$-check corresponds to the edge boundary of a face on the dual lattice, it acts as $X$ on the four qubits surrounding a vertex (which is equal to a face on the dual lattice).
 
\begin{figure}[ht]
\centering
\begin{tabular}{c c}
\includegraphics[width=0.35\textwidth]{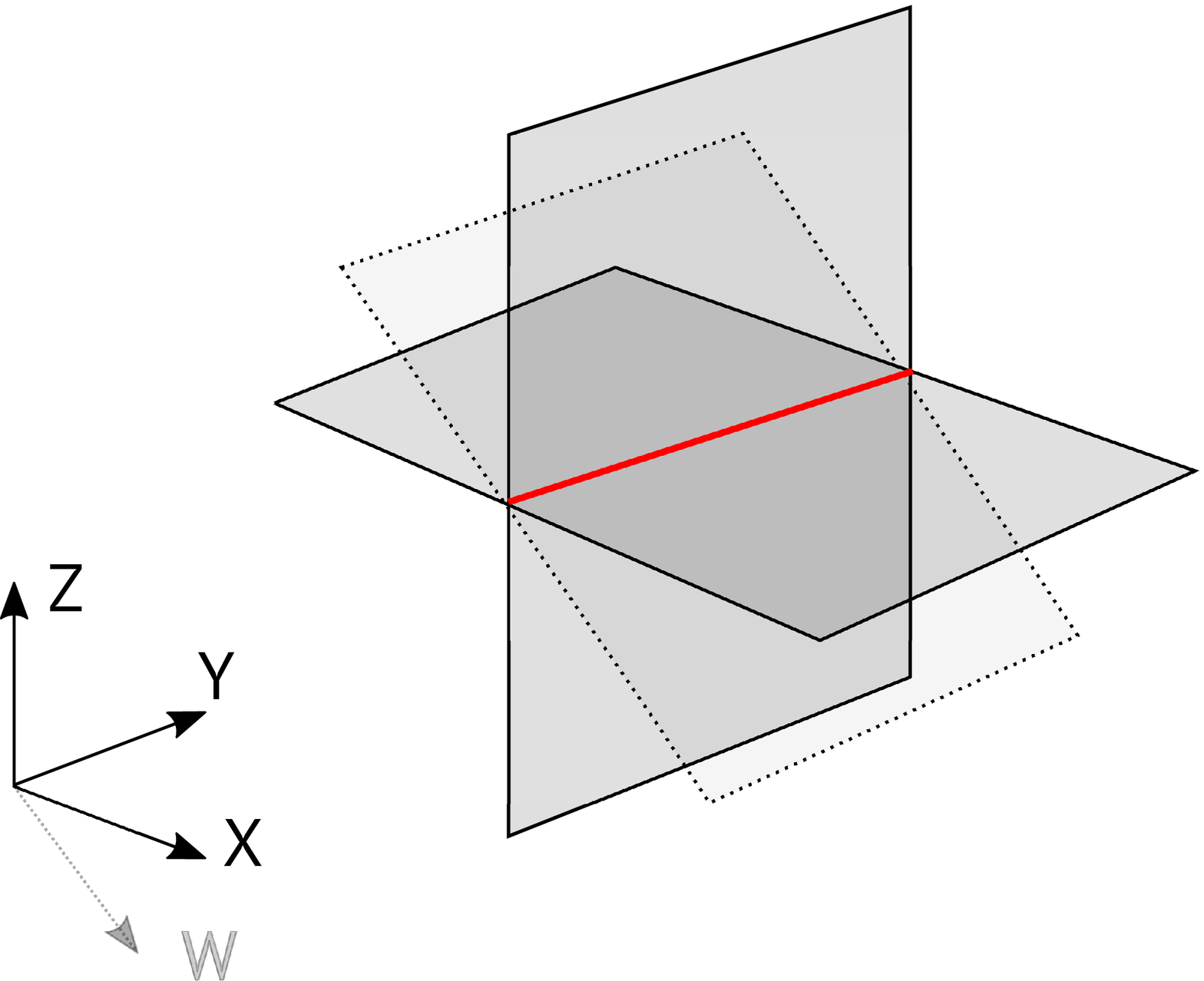} \hspace{2cm} & \includegraphics[width=0.25\textwidth]{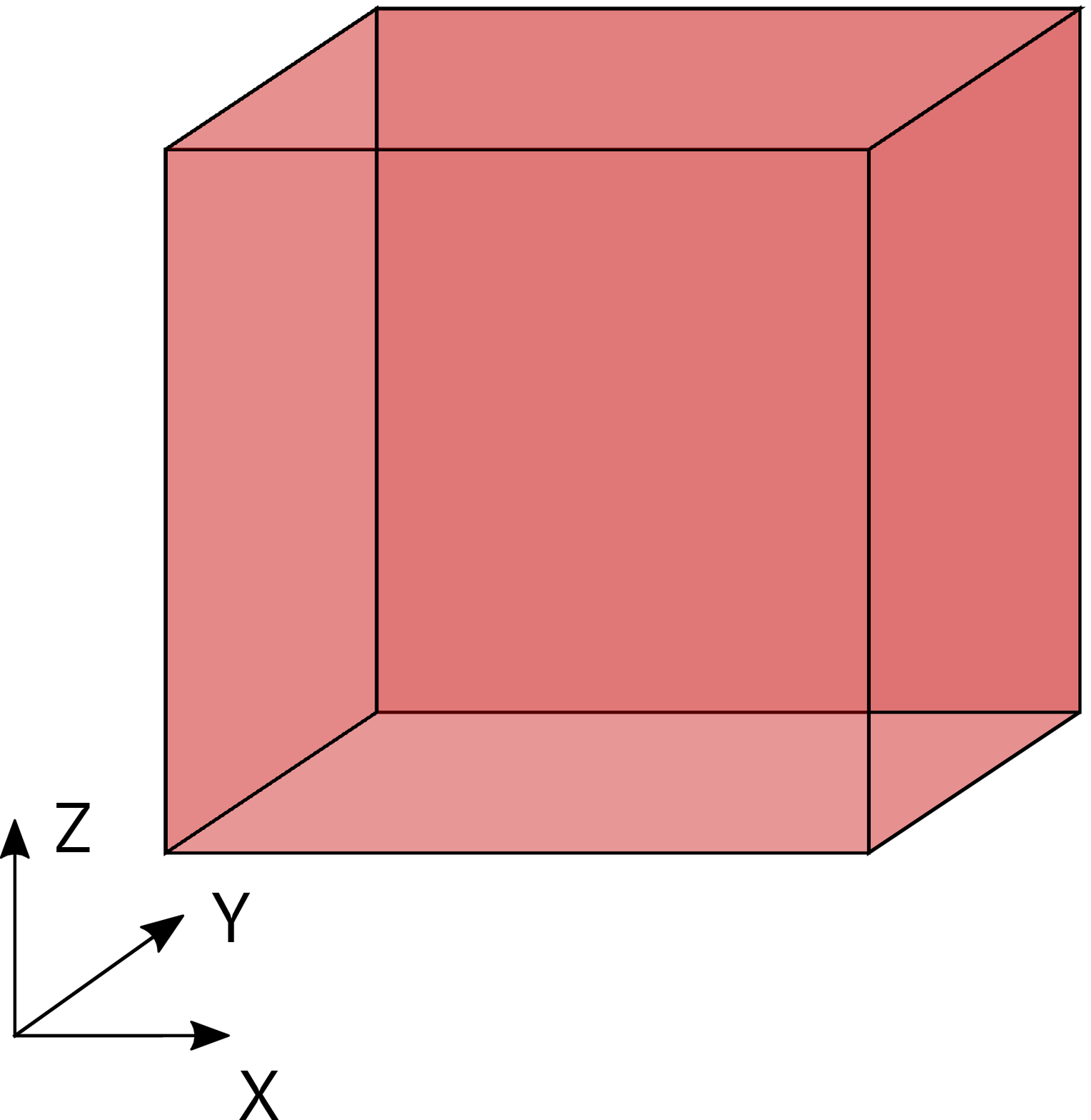}\\
(a) &  (b)
\end{tabular}
\fcaption{(Color Online) Generators of the 4D toric code stabilizer group: (a) With each edge one associates a $X$-parity check acting on the 6 qubits on the faces adjacent to the edge. With each 3D cube one associates a $Z$-parity check acting on the 6 qubits on the faces of the cube. Note that any pair of $X$ and $Z$-checks necessarily overlap on an even number of faces, hence they commute as operators.}
\label{fig:stabilizer}
\end{figure}

For the 4D toric code the parity checks are given in Fig.~\ref{fig:stabilizer}(a) and (b). For the 4D lattice an edge (1-dimensional object) is mapped on a cube on the dual lattice (3-dimensional object) and faces are mapped onto faces. Thus for both codes the action of the $Z$-checks on the lattice is identical to the action of the $X$-checks on the dual lattice. 
For both toric codes the logical operators are pairs of non-commuting operators which act as Pauli-$X$ and Pauli-$Z$ on subsets of qubits. For those operators to commute with the check operators, the sets of qubits on which they act non-trivially should be a line (2D toric code) or a surface (4D toric code) without a boundary in the lattice (for the logical $Z$ operators) or in the dual lattice (for logical $X$ operator). 

On the other hand, a non-trivial logical, say, $Z$ operator cannot be a loop (2D) or a surface without boundary (4D) that encloses a region of the lattice since it would then be a product of all parity $Z$-checks corresponding to faces or cubes in that enclosed region and thus act trivially on the code space. Thus the support of the logical operator should correspond to a non-trivial closed loop for the 2D toric code (resp. non-trivial surface without boundary for the 4D toric code). The parameters of the 2D toric code are $[[2L^2,2,L]]$, i.e. two logical qubits are encoded and their logical operators are non-trivial loops of length $L$ over the 2-torus. The parameters of the 4D toric code are $[[6L^4,6,L^2]]$, i.e. the code encodes six logical qubits whose logical operators correspond to non-trivial surfaces (each a 2D torus).

In one step of quantum error correction (QEC cycle) all parity checks are measured simulataneously: their measured eigenvalues are referred to as the syndrome. In the absence of errors all eigenvalues are $+1$ and we will refer to parity checks with $-1$ eigenvalues as {\em defects} or non-trivial syndromes. 

\begin{figure}[h]
\centering
\begin{tabular}{c c}
\includegraphics[width=0.3\textwidth]{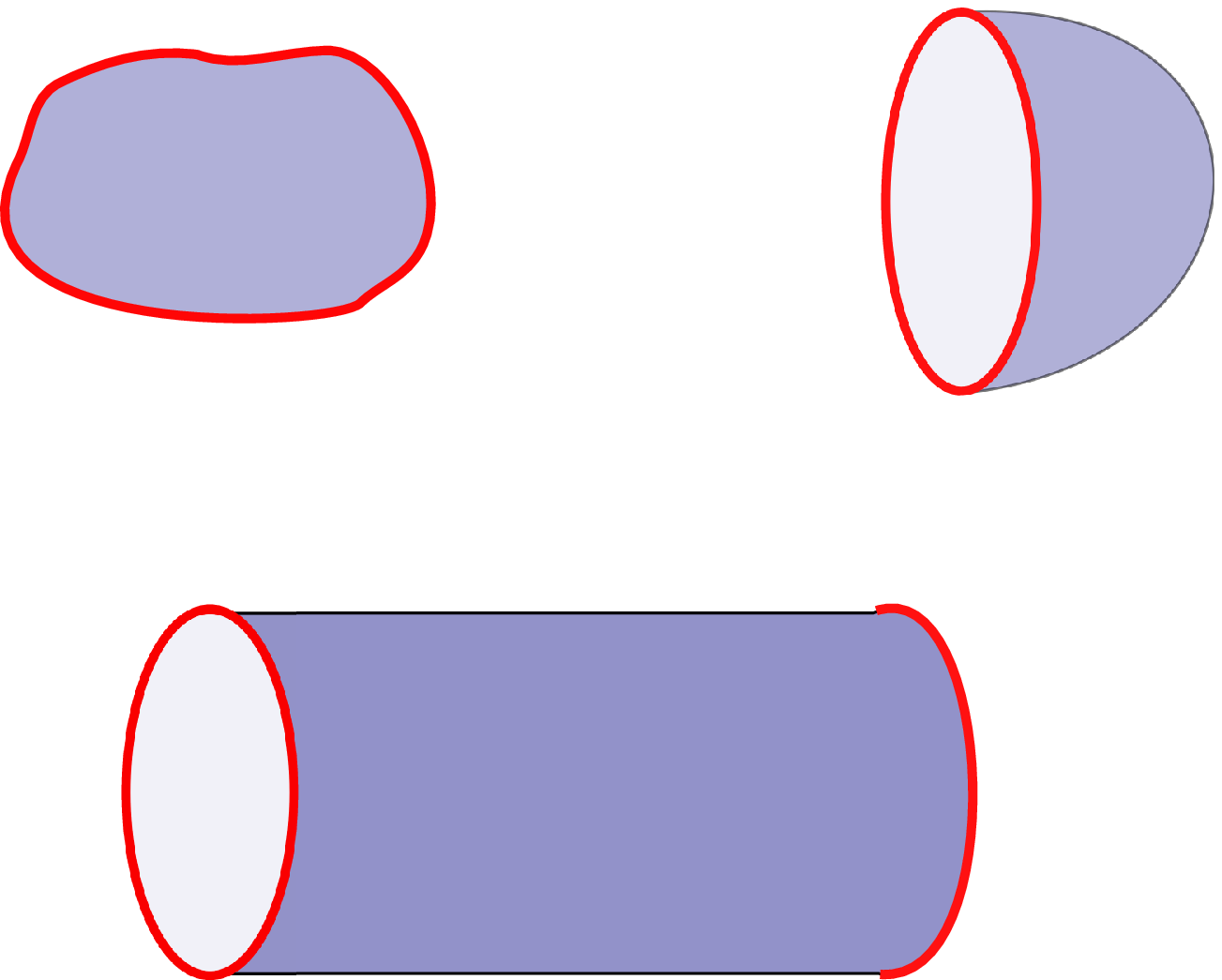} \hspace{2cm} & \includegraphics[width=0.3\textwidth]{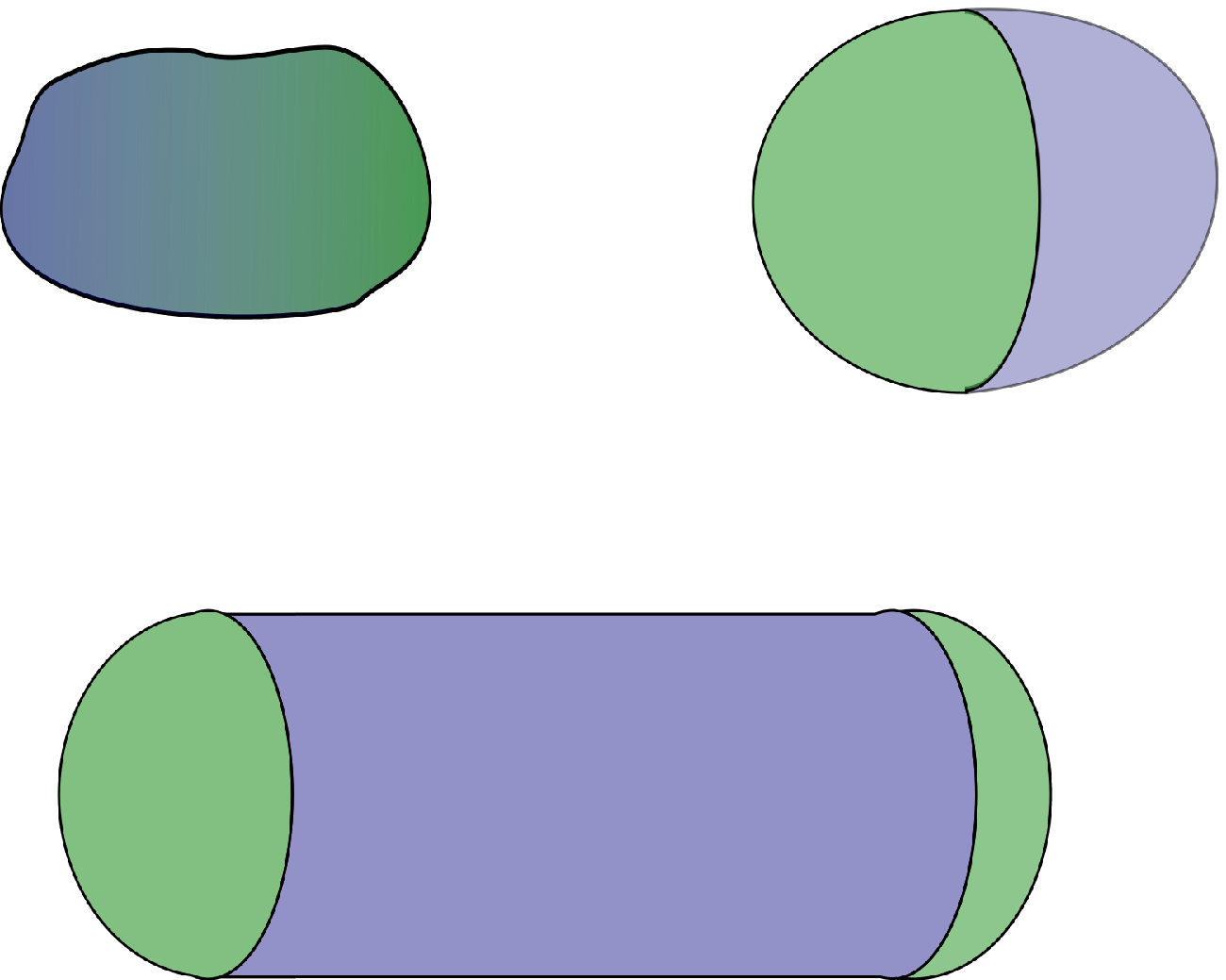}\\
(a) & (b)
\end{tabular}
\fcaption{(a) Errors $E$ (blue) appear as a collection of sheets. The syndrome $S$ (red) is given by the boundary of the errors, just as for the 2D toric code. The syndrome is always a collection of closed loops. (b) An error recovery $R$ (green) applied to the error. The recovery may coincide with the error (upper left) but it can be different in general. The boundary of error and recovery coincide.}
\label{fig:4Dloops}
\end{figure}

The 4D toric code has a feature which is absent from any 2D stabilizer code and which plays a crucial role in error correction and decoding. Namely, the product of the 8 parity $X$-checks corresponding to the edges attached to a single vertex is $I$ (Similarly, the product of 8 $Z$-check cube operators forming the boundary of a hypercube is $I$.) This implies that a string of $X$-checks with $-1$ eigenvalue, i.e. a string of defects, cannot terminate, but instead forms a closed loop. Given a set of $Z$ errors occurring on a set of neighboring faces, the set of $X$-checks anti-commuting with these $Z$ errors (the defects) corresponds to the one-dimensional boundary of this set, see Fig.~\ref{fig:4Dloops}. We will sometimes refer to such connected sets of faces as \textit{error sheets}. For the 2D toric code the defects are points forming the 0-dimensional boundary of {\em error strings}. For the 4D toric code errors can thus be removed by shrinking the length of closed loops locally while no such local error removal rule exists for the 2D toric code \cite{DKLP}.

\subsection{Error Model and Logical Failure}

\begin{figure}[htb]
\centering
\begin{tabular}{c c}
\includegraphics[width=0.3\textwidth]{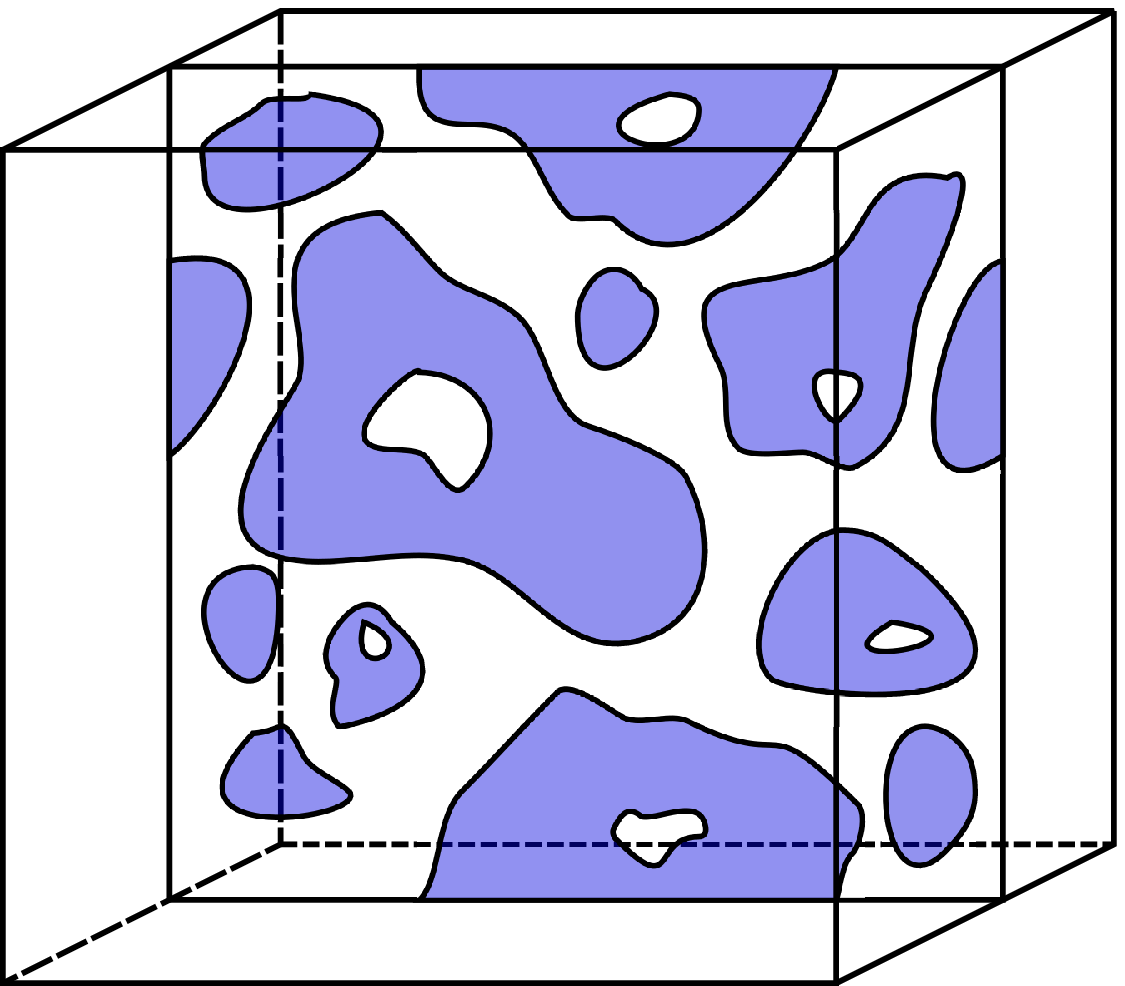} \hspace{2cm} & \includegraphics[width=0.3\textwidth]{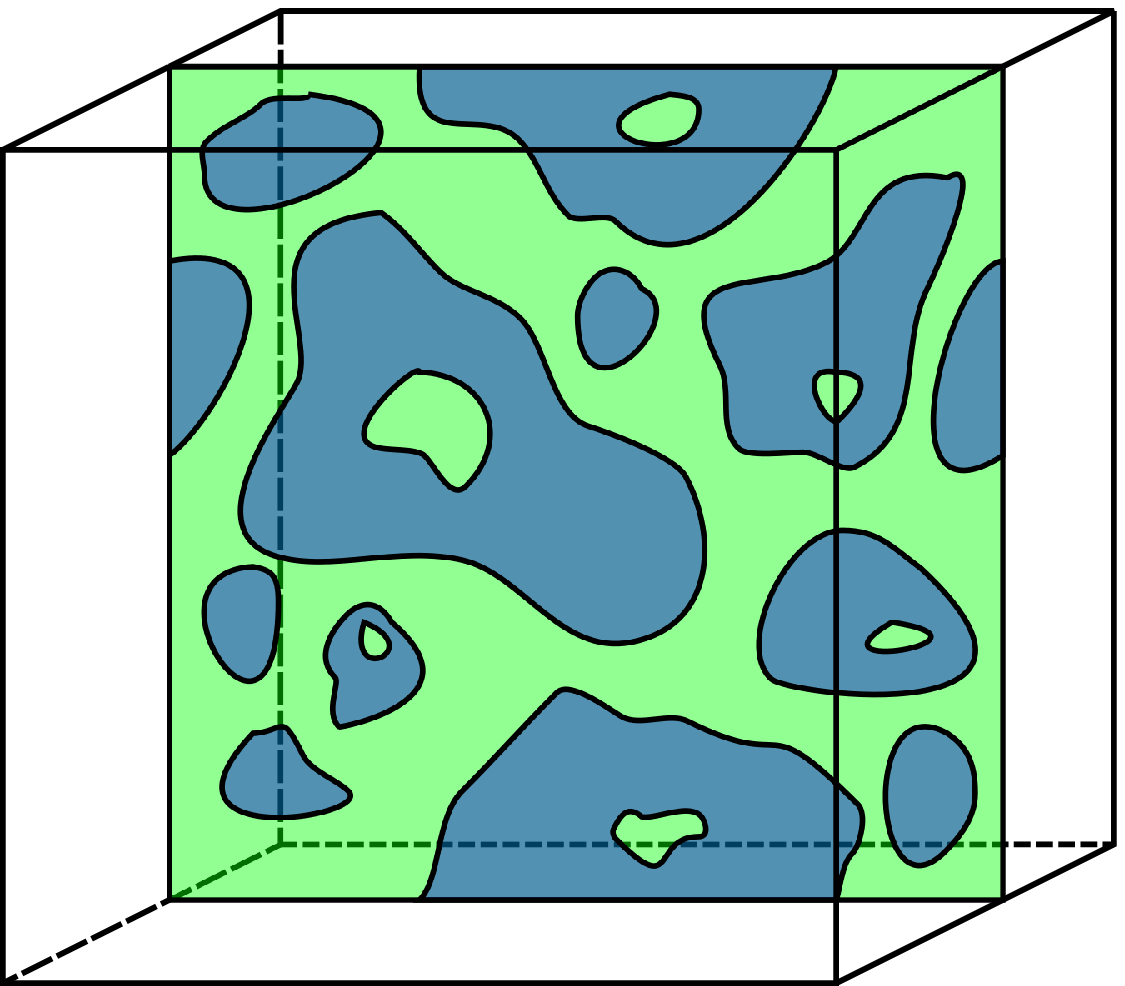}\\
(a) & (b)
\end{tabular}
\fcaption{(Color Online) Example of failed error correction for the 4D toric code. (a) Shown is a 2D slice in a 3D slice of the full 4D lattice. Opposite boundaries are identified. In blue is a `critical' high-weight error $E$ which may lead to failure. (b) Failed recovery $R$ (in green) of the error $E$ (blue). Together they form a homologically non-trivial sheet, i.e. a logical operator.}
\label{fig:failed_correction}
\end{figure}

 To analyze the performance of the decoders we consider an independent $X$ and $Z$ error model: both $Z$ and $X$ errors occur for each qubit independently with probability $p$ in every QEC cycle. Hence a $Y$ error occurs with probability $p^2$ and no error occurs with probability $(1-p)^2$. Given the fact that $X$-checks acts as $Z$-checks on the dual lattice for both codes, we may consider the decoding problem for only, say, $Z$ errors. The decoding and correction problem of $X$ errors will be completely identical. 
 We thus aim to correct for $Z$ errors using the parity $X$-check eigenvalues.  We model a faulty parity check measurement by a perfect check measurement followed by a bit-flip channel which flips the eigenvalue with probability $q$, taking $q=p$ in our simulations. This error model is often referred to as phenomenological noise. After each QEC cycle we perform a recovery step based on the (possibly faulty) syndrome information consisting of $Z$ (phase) flips on a set of qubits. We assume that this correction is implemented noiselessly. In practice, it is known that the correction need not be applied but can be kept in software as a Pauli frame \cite{FMMC:review} warranting the assumption of being noiseless. 
 
The general strategy of the decoders is to either move the defects towards each other to annihilate (2D toric code) or to minimize the length of the defect loops (4D toric code). After each recovery step we perform a logical failure test in our simulations. During this failure test we check if we would be able to retrieve the logical quantum data were we to know the syndrome with 100\% confidence (no measurement errors). Logical failure then occurs when the real error $E$ on the data together with the inferred recovery $R$ form any of the logical operators of the code. When the syndrome record is error-free, $E+R$ will have no boundary (0-dimensional in 2D and 1-dimensional in 4D), see e.g. Fig.~\ref{fig:4Dloops} for the 4D case. In Fig.~\ref{fig:failed_correction} we show an example in which logical failure occurs for the 4D toric code as $E+R$ forms a logical operator. 

We report on the (average) memory time as the number of rounds of this procedure we can run on average until logical failure. If error correction is effective this memory time increases with $L$. If error correction is ineffective, the memory times decreases with $L$. We will refer to the noise threshold as the probability $p$ (for $q=0$ or $q=p$) where the cross-over from ineffective to effective quantum error correction occurs. For some of the plots, e.g. Fig.~\ref{im:rough_study} and Fig.~\ref{fig:DKLP_results}, this cross-over point can depend on $L$ (as it is receding with $L$) but one expects that the sequence of cross-over points converges to a single point for sufficiently large $L$. We do not claim that the existence of a threshold has been proven for all decoders that we consider.

\subsection{Previous Work on Decoders and Noise Thresholds}

\subsubsection{2D Toric Code}
For noiseless parity checks ($q=0$) the 2D toric code has been found to have a noise threshold of $p_c\approx 10.3\%$ using the minimum weight matching (MWM) decoder \cite{WHP:threshold}. This is close to the optimal possible threshold of $10.9\%$ for this code (the optimum is set by the intersection of the phase boundary in the corresponding 2D random bond Ising model with the so-called Nishimori line). For the same error model a renormalization group (RG) decoder \cite{DP:fast} gives $p_c \approx 9\%$. For noisy parity check measurements with $q=p$, the 2D toric code threshold using MWM goes down to $2.93\%$ \cite{WHP:threshold}, while the RG decoder gives $p_c \approx 1.9\%$ \cite{DP:3Ddecoding}.  This RG decoder can be viewed as a hierarchical decoder such as the Harrington decoder but with arbitrarily high classical speed so that nonlocal computation is instantaneous.

Many other decoders exist for the 2D toric code.  For example, Ref.~\cite{BH_RGdecoder} introduces a RG decoder for general topological stabilizer codes where clusters of errors are identified and removed at different length scales.
Other decoders strive to represent a local physical system coupled to the toric code such that the decoding process can be interpreted as passive dissipative mechanism stabilizing the toric code, see e.g.  \cite{2009PhRvB..79x5122H, Herold2015, Herold2015, fujii+:dissipative}. The Harrington decoder is not strictly local in the sense that the memory for some cells scales logarithmically with the linear system size $L$. Such scaling also occurs in systems obtained by coupling the toric code to an auxiliary system \cite{2009PhRvB..79x5122H, Herold2015} requiring an interaction strength or precision of the interaction scaling with system size. 

We note that the estimated numerical noise threshold of the Harrington decoder in \cite{Herold_other} of $p_c \approx 0.001\%$ is far below the results in this paper. The Harrington decoder has also been generalized to work for codes with non-Abelian anyons \cite{DP_nonAbelian} although the implementation in \cite{DP_nonAbelian} lets the computation at higher less-local levels in the hierarchy become non-local, thus incurring no communication delays.

\subsubsection{4D Toric Code}
For noiseless parity checks ($q=0$), the 4D toric code has been conjectured in Ref.~\cite{takeda} to have the same maximal threshold of $10.9\%$ as the 2D toric code. Ref.~\cite{takeda} maps the 4D toric code to a statistical physics model called the 4D random plaquette gauge model which they argue to have the same phase diagram as the 2D random bond Ising model. The conjecture is supported by numerical simulations reported in \cite{arakawa2005self} where the authors find a threshold of $10.11 \%\pm 0.02 \%$. It is quite possible that this optimal threshold also holds for noisy parity checks ($q=p$). Noisy parity check information can be repaired using minimum weight matching of vertices in the 4D hypercubic lattice \footnote{This procedure will fail when end-points of syndromes are matched so that a closed non-trivial defect loop results. A single homologically non-trivial defect loop is not a valid syndrome as it is not the boundary of a surface. One thus expects that below the threshold of the 4D random bond Ising model this repair process can be succesful and will induce only local errors.}. A rigorous argument showing that the threshold for noisy parity checks is the same as for noiseless parity checks would show that the 4D toric code is very robust against errors.

The computational complexity of minimum weight decoding for the 4D toric code is not known \footnote{We can reformulate the decoding problem as a special case of finding the minimal $T$-join in a hypergraph $G=(V,E)$. The vertices $V$ represent $X$-check operators and let the non-trivial syndrome set be $T \subset V$. Each hyperedge $e \in E$ corresponds to a qubit (thus each hyperedge acts on 4 vertices in the 4D hypercubic lattice). Let $\delta(v)$ be the set of hyperedges that have $v$ as boundary: for the 4D hypercubic lattice these are the 6 faces touching each edge as in Fig.~\ref{fig:stabilizer}(a). Let $x \in \{0,1\}^{|E|}$ be an incidence vector with 0,1 entries and let $x(S)$ with $S \in E$ be defined as $\sum_{i\in S} x_i$. A $T$-join $J \subset E$ is a subset of hyperedges (given by incidence vector $x_i=1, \forall i\in J$ and $x_i=0, \forall i\notin J$), such that for each vertex $v$ in $T$, $x(\delta(v))$ is odd, while for all $v \in G\backslash T$, $x(\delta(v))$ is even. A minimal $T$-join minimizes $|J|$ and thus corresponds to the minimal weight $Z$ error which leads to the $X$-syndrome. The general hypergraph $T$-join problem is NP-complete. The $T$-join problem for graphs is solved by minimum weight matching, see \cite{book:schrijver}.}. As mentioned, computationally-efficient minimum weight matching can be used to convert the syndrome record with defect strings with boundaries to a set of closed loops. Given a set of closed loops, the minimum weight decoding problem is to find a surface of errors whose one-dimensional boundaries are the given loops. It is not known whether there exists an efficient algorithm for this problem for the 4D hypercubic lattice.

Ref.~\cite{DKLP} has suggested two local (cellular automata) decoding procedures for the 4D toric code. The first procedure which we refer to as the DKLP rule is the following. Given a syndrome we choose a set of faces $F$ where no two faces share an edge. For any face $p \in F$ we apply a $Z$-flip  with 
a) probability 1 if three or more of its parity $X$-check edges are defects, b) with probability $1/2$ if exactly two of its edges are defects and c) we do not perform a flip if one or none of its edges are defects. This procedure can only decrease or maintain the number of defects, leading thus to the shrinking of error sheets.

A similar local rule for shrinking error sheets can be formulated using a 4D version of Toom's NEC rule \cite{toom, grinstein2004}. Toom's rule for the 2D Ising model consists of anisotropically shrinking error regions by flipping a spin iff its parity with both its northern neighbor and eastern neighbor is odd.
In \cite{PCC:local, PastawskiPhD} the authors have implemented a 4D version of this local deterministic rule obtaining a noise threshold of the error model in this paper of $p_c \approx 3.75\%$ at $q=0$ (in this simulation they apply the CA rule many times after obtaining the syndrome record). In Section \ref{sec:toom_dklp} we re-analyze the DKLP rule and Toom rule in our setting and compare their 
performance with a local decoder introduced by Hastings \cite{hastings}. Hastings' decoder was formulated to argue that 4D hyperbolic surface codes, first described in \cite{guth_lubotzky}, can be efficiently decoded by local rules. We describe our implementation of this decoder in Section \ref{sec:hastings}.

\section{Harrington Decoder for 2D Toric Code}
\label{sec:harrington}

\begin{figure}[htb]
\centering
\includegraphics[width=0.8\textwidth]{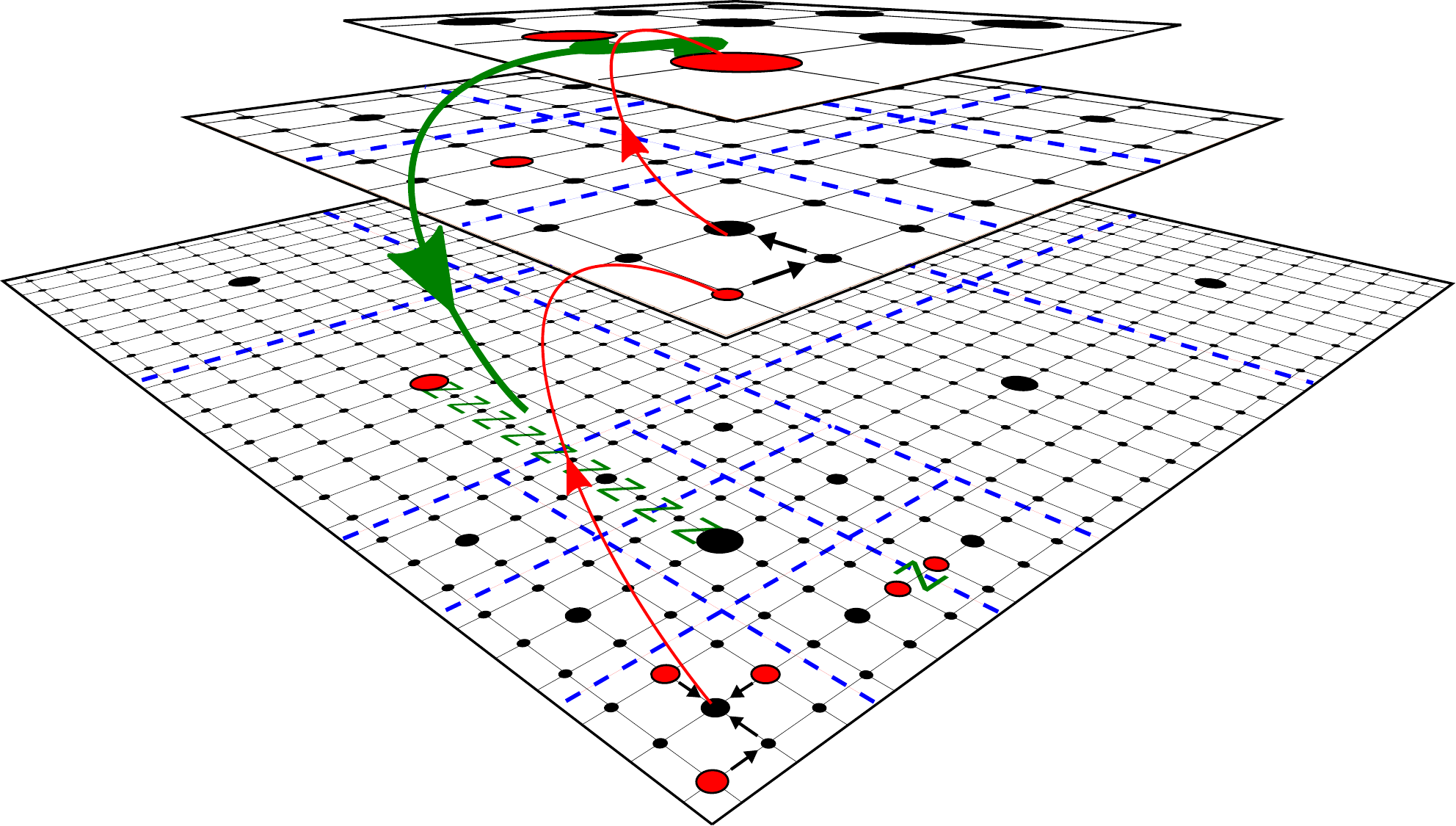}
\fcaption{\label{fig:harrington} (Color Online) Overview of the Harrington decoder. The lower grid represents a toric code lattice ($L=27$) with qubits on the edges. The black dots represent locations of parity $X$-checks and 0-cells. Groups of $3\times 3$ cells form a colony indicated by the dotted blue lines. The slightly larger black dots are the representatives of the 1-cells, 2-cells etc. located at the center of these colonies.  The CA rules are such that defects are moved towards the center of the colony. For example, the three defects in the forefront, indicated by red dots, are moved towards the center of the corresponding colony, indicated by the arrows. Defects in neighboring colonies can be directly annihilated in pairs, indicated by the single green $Z$ between two other defects. If defects are not annihilated by being moved to the center of the colony, the remaining defect, located at the center of the colony can now be moved to the next layer in the hierarchy structure, indicated by the middle grid. Each black dot in this middle grid now represents a 1-cell. The 1-cells are organized into groups of $3\times 3$ forming a colony, again indicated by the dotted blue lines. The 1-cell obtains a 1-syndrome which is locally calculated from the defects located at its representative 0-cell, indicated by the curved red arrows.  Again, the defects at level-1 are moved to the center of the colony. Remaining defects are annihilated in pairs in the next, and last, level of the hierarchy, indicated by the top grid. Here, each black dot is a 2-cell. The annihilation of defects on higher levels in the hierarchy is eventually communicated and executed locally, i.e. by the 0-cells. This is indicated by the green curved arrow and the string of Pauli $Z$ operators, annihilating the defects appearing at the 2-cells. } 
\label{fig:harrington_pic}
\end{figure}

In this section we review the decoder introduced by Harrington in his PhD thesis \cite{thesis:harrington}. We start by giving a high level overview of the decoder. The decoder is a cellular automaton consisting of $L^2$ cells, one for each parity $X$-check, see Fig.~\ref{fig:harrington_pic}. Functional groups of $Q\times Q$ cells are called \textit{colonies}, groups of $Q\times Q$ colonies can be called \textit{super-colonies} and so on. This induces a hierarchy of groupings until the whole lattice is included, hence $L=Q^k$ for some integer $k$ \footnote{In principle, colony sizes at different levels of the hierarchy could be different. One can also choose $L > Q^k$ so that part of the lattice is covered by a supercolony of size $Q^k \times Q^k$ while the rest is covered by smaller colonies and possibly varying $Q$.}. 

Each cell in the automaton, also referred to as 0-cell, is defined by its state (memory) and a transition function. The transition function describes how the state of a cell evolves after each time step, possibly depending on the state of its eight neighbors as well the value of the syndrome at the corresponding $X$-check. In principle, at each time step or QEC cycle, a 0-cell receives a new parity check record. In Section \ref{sec:speed} we consider how well the decoder functions when the rules of the CA can be executed for more time steps in between QEC cycles.

The transition function of a cell describes when to perform a $Z$-flip on one of the four qubits on which the corresponding $X$-check has support. Each colony functions itself as a cell at a higher, less-local level, implementing error correction at a larger length scale. A colony of size $Q \times Q$ thus functions as a 1-cell, receiving a 1-syndrome and executing error correction at level 1. The memory and computation of a $1$-cell is located at the 0-cell at the center of the colony. We will call this 0-cell the representative of the 1-cell. Similarly, the memory and computation of a $2$-cell, which performs error correction at level-2 using a $2$-syndrome, is located at the 0-cell residing at the center of a colony of colonies etc., see Fig.~\ref{fig:harrington_pic}. In order for the higher-level cells to execute the same rules as the 0-cells, explicit communication between the physical representatives of the $i$-cells is needed. Time lags due to this communication which takes place via the $0$-cells are explicitly part of the functioning of the decoder. 

The transition function of $0$-cells is programmed to reduce the number of defects either by locally annihilating them in pairs, or by moving them to the 0-cell at the center of a colony \footnote{The center of the colony is unambiguous when $Q$ is odd, otherwise one just makes a consistent choice between the four center $0$-cells.}. This can result into defects being stuck in the center of colonies. This in turn will set a non-trivial $1$-syndrome which should be removed by error correction at level-1 or higher. The $1$-syndrome is not determined every time step, but after a work period $U$ and in a manner which is robust against syndrome and qubit errors. Error correction at level-1 thus occurs $U$ times slower than error correction at level 0. Similarly, the $2$-syndrome is obtained every $U^2$ steps equal to a level-2 work period etc.

In the next Section we discuss the functioning of $0$-cells in a colony in some more detail. In Section \ref{sec:higher} we explain how the syndromes are obtained and rules are executed at higher levels of the hierarchy.

\subsection{Effect of Local Update Rules}
Per time step (QEC cycle) a 0-cell performs a $Z$-flip on at most a single qubit and only when the 0-cell has a defect. Which qubit is flipped depends on the relative location of the 0-cell in the colony and the presence of defects in its neighborhood. If a defect is present at one (and only one) of the four nearest-neighbor 0-cells (on horizontal and vertical edges), the CA is programmed to flip the qubit between the two defects. This is implemented by only one of the two 0-cells to prevent double flipping. The other 0-cell will not perform any $Z$-flip. A certain preference ordering of the four nearest-neighbors is implemented and used if more than one of these neighbors have defects, see Appendix A. If no defects are present at its four direct neighbors, the 0-cell will continue by checking for defects at its four next-nearest-neighbor 0-cells and will similarly try to annihilate a pair of defects by performing a $Z$-flip on the qubit between them. We refer to Appendix A for details. As a last step, if the nearest-neighbor nor the next-nearest-neighbor 0-cells have no defects, the 0-cell will perform a $Z$-flip in order to move the defect towards the center of the colony. This last step plays an important role when the syndrome record is noisy and single defects can occur. The CA rules ensure that a single defect leads to a $Z$-flip. In the next QEC cycle, assuming no syndrome error, this $Z$-flip will then be corrected.

\subsection{Error Correction at Higher Levels}
\label{sec:higher}
The behavior of the 0-cells is implemented identically at higher levels by $i$-cells which are physically located/represented at centers of colonies. We discuss the following four functionalities:
\begin{enumerate}
\item An $i$-cell should be able to get an $i$-syndrome and decide whether it is either a defect or not. 
\item An $i$-cell should be able to learn if neighboring $i$-cells have $i$-syndromes which are defects. 
\item Based on this information an $i$-cell should decide where to move its defect: this is done using the same update rules as described in the previous Section and the Appendix.
\item An $i$-cell should be able to actually move its defect, that is, apply $Z$-flips on physical qubits.
\end{enumerate}
The second (ii) and fourth (iv) steps require communication at a length scale $Q^i$ while communication between adjacent $0$-cells is assumed to occur in a single time step/QEC cycle. This communication is implemented by using 0-cells which lie on a straight line between the representatives of the two $i$-cells. In the second step (ii) the communication also takes place between $i$-cells which are diagonal neighbors. Since communication between $i$-cells takes $Q^i$ time steps the notion of an $i$-level work period, being $U^i$ time steps ($U\geq Q$), is introduced. During this work period, $U^i$ time steps, the $i$-cells evolve as proper cells do for 1 time step.\\

(i) Deciding whether an $i$-cell has a defect is executed each work period, i.e. $U^i$ time steps. The following averaging procedure decides whether the $i$-cell has a defect. In a $Q\times Q$ colony of $i-1$ cells, $i-1$ cells update their defects every $U^{i-1}$ timesteps and drive the defects to the center $i-1$ cell. The presence of a defect at level-$i$ is then determined from the stored record of $U$ $i-1$-defects at the center $i-1$-cell. Harrington prescribes to bin this $U$-tuple into $b$ blocks of length $b$ ($U=b^2$). If during at least $f_cb$ of the $b$ blocks ($0<f_c<1$), the $i-1$-cell has a defect for at least $f_cb$ of the time, the $i$-cell concludes that it also has a defect. This procedure not only builds in robustness against measurement errors, it also plays a role in Harrington's proof that the decoder has a threshold (for $b\geq20$). The downside of this procedure is that it gives rise to large values for $U$. We observed numerically that using a simplified decision procedure, namely that an $i$-cell has a defect iff its center $i-1$-cell has a defect for at least $f_cU$ times, gives rise to longer memory times, see Section \ref{sec:numerics}. \\

(ii) Similarly, in the second step, an $i$-cell should learn whether a neighboring $i$-cell has a defect. During a work period the representative of the $i$-cell obtains $U$ independent defect values from the representative of a neighboring $i$-cell. In other words, each $i-1$-syndrome bit of a center $i-1$-cell is `broadcasted' to neighboring (diagonally and horizontally/vertically) center $i-1$-cells. An alternative is to first let each $i$-cell compute its $i$-syndromes and only then commmunicate it to its neighboring $i$-cells. Such sequential processing would lead to larger delays while letting the $i-1$-center cells stream their data and letting the $i$-cells compute the neighboring $i$-syndromes from this data themselves is more time-efficient. One should note that due to communication lag a fraction of these defect values originate from the previous work period but for $U \gg Q$ this is a small effect.
To determine the $i$-syndrome of a neighboring cell on the basis of this stream, one of the two procedures above can be used, but Harrington allows some flexibility in choosing a different threshold parameter $f_n \neq f_c$. 
By taking $f_n\neq f_c$, it may occur that cells take different conclusions. Some $i$-cell concludes that a neighboring $i$-cell has a defect, while the particular neighboring $i$-cell concludes it does not, or vice-versa. As Harrington explains, $f_c$ should be large and $f_n$ small for good performance. Large $f_c$ (high threshold for level-1 defects) makes it likely that level-1 errors correspond to real error chains between 1-cells. Large $f_c$ also increases the probability of not dealing defect, making defects less mobile, which may suppress performance at higher error rates. Small $f_n$ on the other hand reduces the probability of wrongly moving a defect occurring in an $i$-cell to the center of an $i+1$-cell, instead of pairing it with defect in a neighboring $i$-cell. In proving a threshold for his decoder Harrington uses $f_n = 1-f_c = \frac{4}{5}$.\\

(iv) In the last step the $i$-cells should communicate back to the $0$-cells to flip a chain of qubits running between their representatives (since this is where the errors have been moved to). The communication takes $Q^i$ time steps, the flipping of the qubits itself is done simultaneously at the end of a work period. 
The communication delay means that qubits are corrected with some delay. A single string of errors between two $i$-cells representatives which appears late in a work period 1 will set the $i$-syndrome to be a defect only in the next work period 2, so that at the end of work period 3 the actual correction will take place.
An important aspect of the implementation is that the decoder adjusts the incoming syndrome record for the corrections upon which a decision has been taken. In other words, if an $i$-cell decides that a string of $Z$-flips has to take place, it takes time for this to be executed. During this time the syndrome continues to indicate the need for correction. Without taking this into account the $i$-cell may decide twice for the same correction leading to unwanted flips of qubits.

The C$^{++}$ code for the Harrington decoder can be downloaded from GitHub, \url{https://github.com/kduivenvoorden/LocalToricDecoder} where one can also find a Mathematica script with which one can generate a movie based on the output data of the decoder so that one can see the decoder in action. An example of such a movie including a description of the decoder can be found at \url{https://youtu.be/WKos_vcY3bI}.

\section{Numerical Results with the Harrington Decoder}
\label{sec:numerics}

\begin{figure}
\centering
\includegraphics[width=0.5\textwidth]{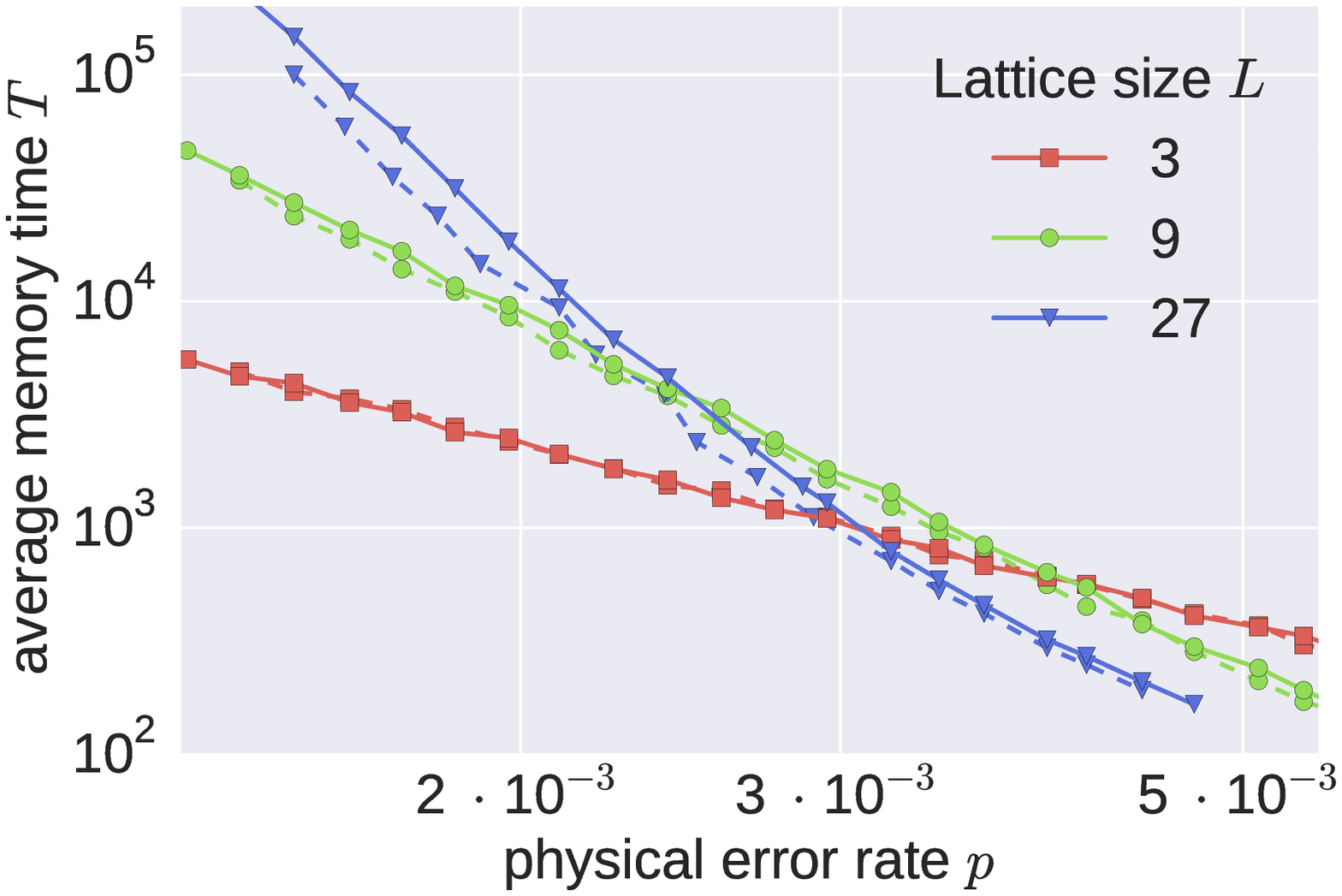}
\fcaption{(Color Online) Average memory times either using (solid lines) or not using (dashed lines) a rough test prior to doing minimal weight matching, using the parameters $Q=3,U=10,f_n=4/10,f_c=9/10$.} 
\label{im:rough_study} 
\end{figure}

After each time step of the cellular automaton, we want to check whether we can recover the quantum data using a global decoder and having perfect knowledge of the syndrome. Unfortunately, running minimum weight matching (MWM) after every simulation step is slow (resulting in lengthy computation when the memory time is long). Therefore we use a rough test (described by
Harrington \cite{thesis:harrington}) which {\em preselects} the cases in which minimum weight matching is run. If the rough test is passed, we call no logical failure. If the rough test fails, we determine logical failure or no logical failure via MWM.  The rough test functions as follows: Count both the number of rows and columns where the number of $Z$ errors is odd. We conclude failure if one of these numbers is higher than $\frac{L}{2}$.  The rough test is neither necessary nor sufficient for a logical error after minimal weight matching \footnote{We give a
counterexample for the rough test being necessary: Consider a $5\times 5$
lattice with defects at the positions $(0,0),(1,2),(2,2),(3,0)$ and errors at positions $(0,.5)(0,1.5),(.5,2),(2.5,2),(3,1.5),(3,.5)$. This error
set passes the rough test, indicating no failure. The MWM decoder will introduce a
logical error.}. Nevertheless a numerical study implies that it is a necessary condition 
for a logical error with high probability, see Fig.~\ref{im:rough_study}.  Note that the rough test could just be used as a final read-out step of the memory. One measures all qubits in the $X$-basis and wants to estimate the eigenvalue of $\overline{X}_1$ and $\overline{X}_2$ of the two encoded qubits. One estimates $\overline{X}_1$ by taking the majority value of all realizations of $\overline{X}_1$ as product of $X$s in a row (and similarly one estimates $\overline{X}_2$ by taking the majority value of all realizations of $\overline{X}_2$ as product of $X$s in a column). Only rows (or columns) with odd numbers of $Z$ flips can change this value and the rough test thus captures whether such read-out method will be succesfull.

\subsection{Determining the optimal parameters $Q,U,f_n,f_c$} 

\begin{figure}
\centering
\begin{tabular}{c c}
\includegraphics[width=0.45\textwidth]{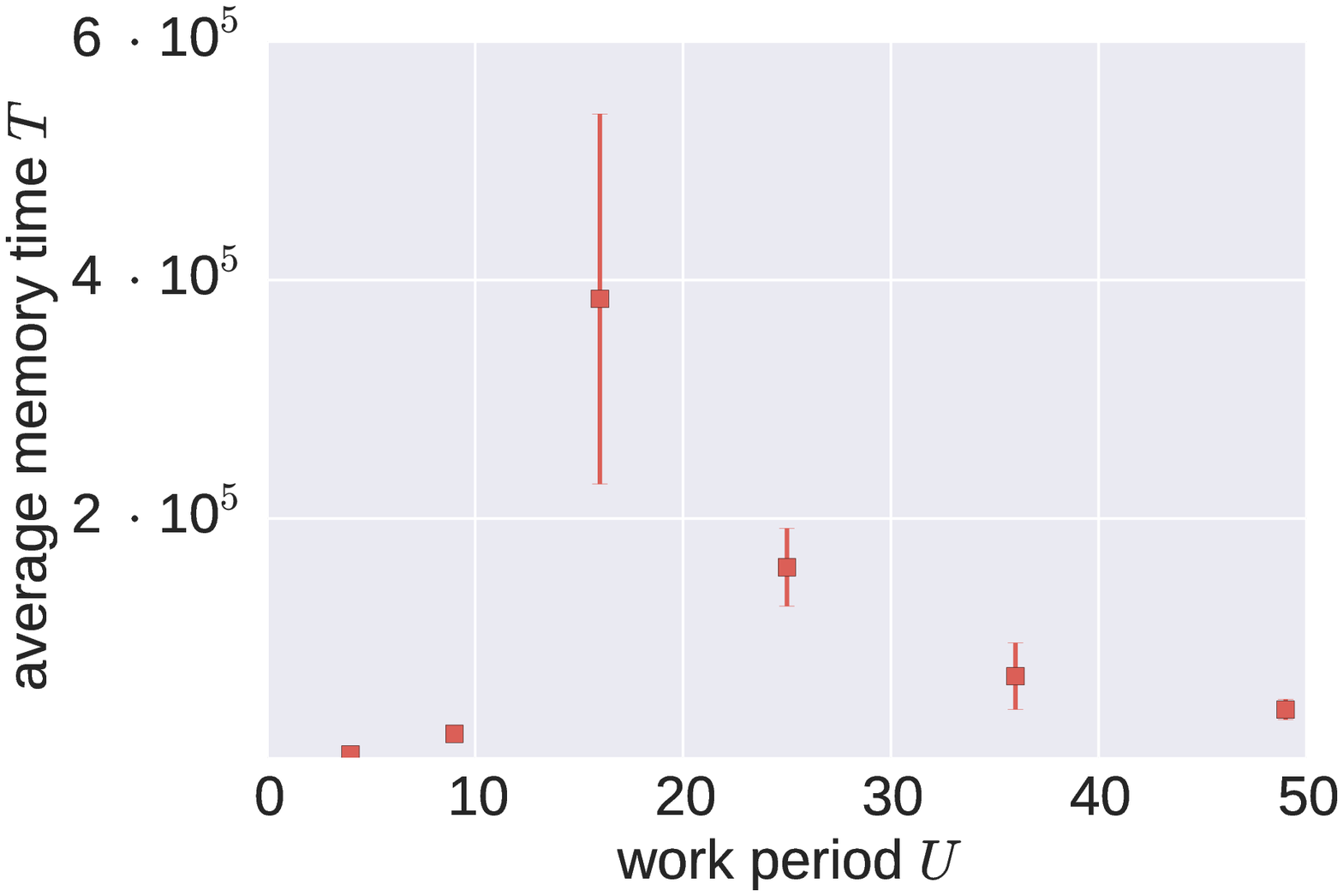} \hspace{1cm} & \includegraphics[width=0.45\textwidth]{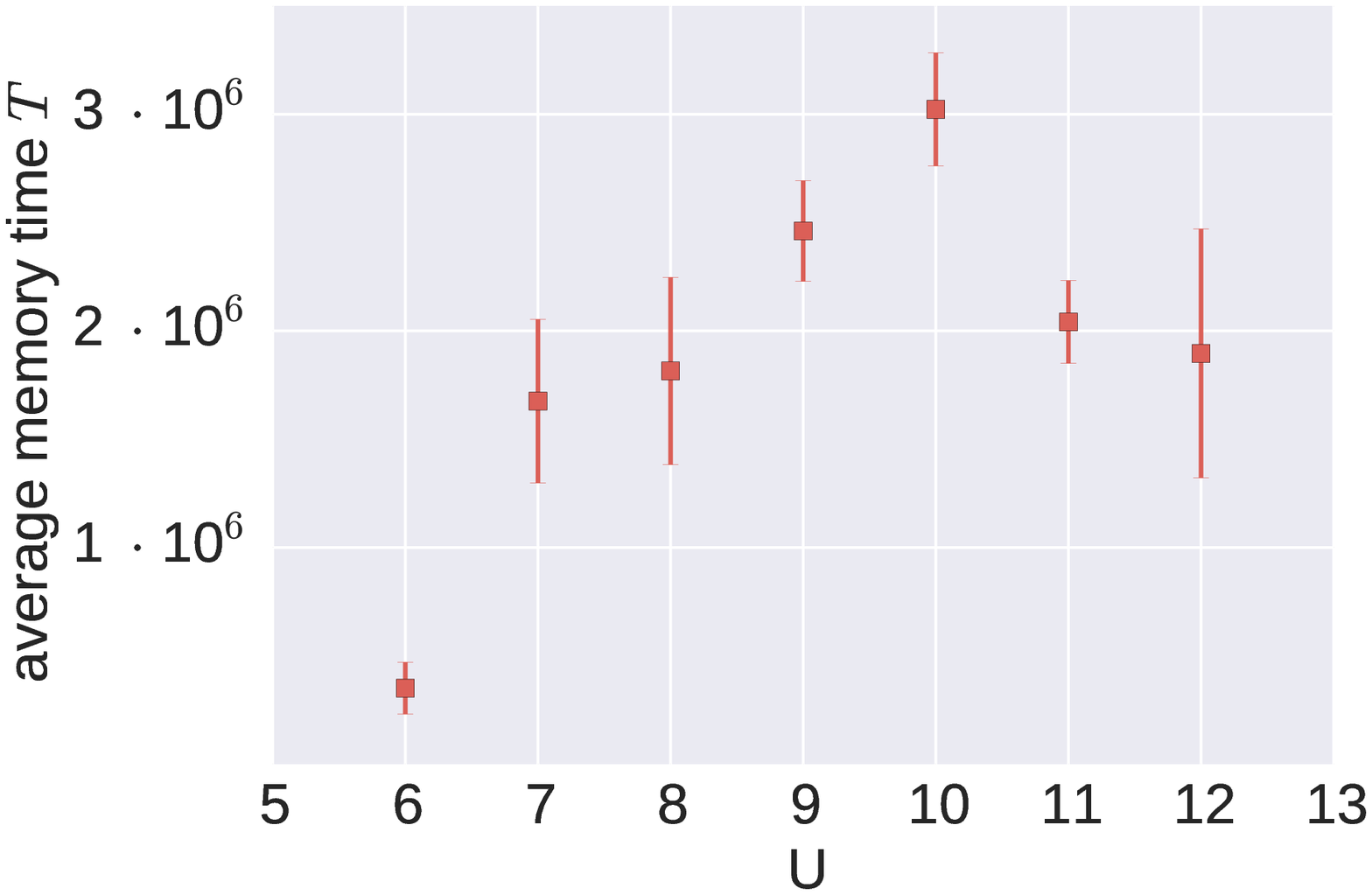}\\
(a) & (b)
\end{tabular}
\fcaption{(Color Online) Average memory time vs. the
work period $U$ with optimal parameters $f_n$ and $f_c$ at $Q=3$ and $L=27$ for the division strategy (a) and non-division strategy (b).}
\label{im:optimal} 
\end{figure}

We investigate whether there is numerical evidence for a threshold for low values of $Q$ and $U$ with the best possible memory times. We have described two strategies for concluding if a defect occurs at higher hierarchy levels, depending on whether one divides $U$ into $b$ blocks of length $b$ or not. We have looked at both strategies. For the division strategy we study the memory time at fixed  colony size $Q=3$, system size $L=27$ and error probability $p=q =0.1\%$ and vary $U\in\{9,16,25,36,49\}$. For each value of $U$ we determine the optimal value for both $f_n$ and $f_c$ and report on the memory time at this optimal value, see Fig.~\ref{im:optimal}(a). The highest memory time is obtained for $U = 16$, $f_n = 1/2$ and $f_c=3/4$. For the non-division strategy we again fix the error probability to  $p=q =0.1\%$ and system size to $L=27$ and vary $U\in[6,12]$. For all configurations we determine the optimal value for both $f_n$ and $f_c$ and report on the memory time at this optimal value, see Fig.~\ref{im:optimal}(b).  The highest memory time is obtained for $U = 10$, with $f_n = 4/10$ and $f_c=9/10$. Also, we observe that the obtained memory time is larger than the largest memory time obtained using the division strategy. For all further simulations we thus use the non-division strategy with $U = 10$, $f_n = 4/10$ and $f_c=9/10$.

\subsection{Threshold}
\begin{figure}[h]
\centering
\begin{tabular}{c c}
\includegraphics[width=0.45\textwidth]{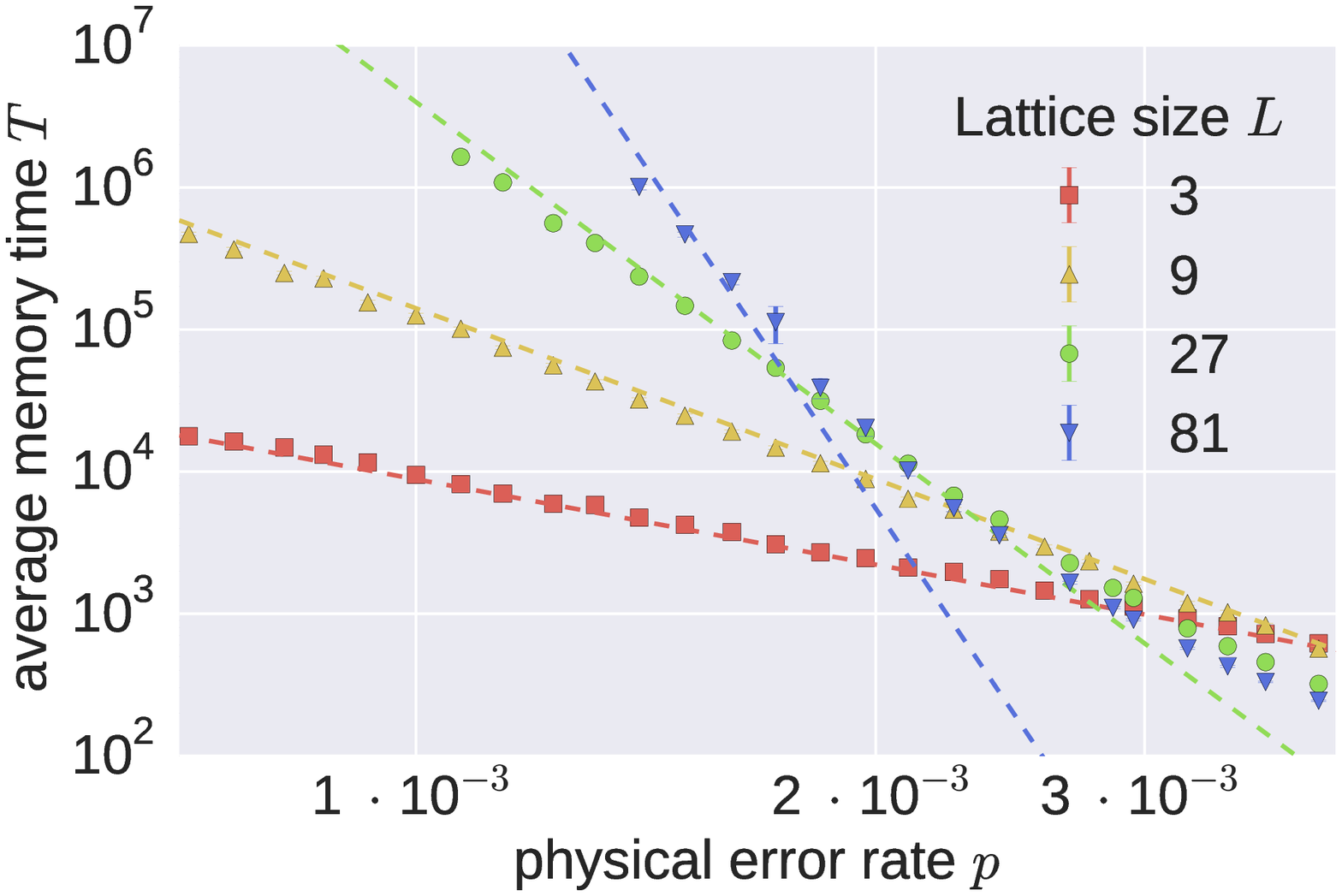} \hspace{1cm} & \includegraphics[width=0.45\textwidth]{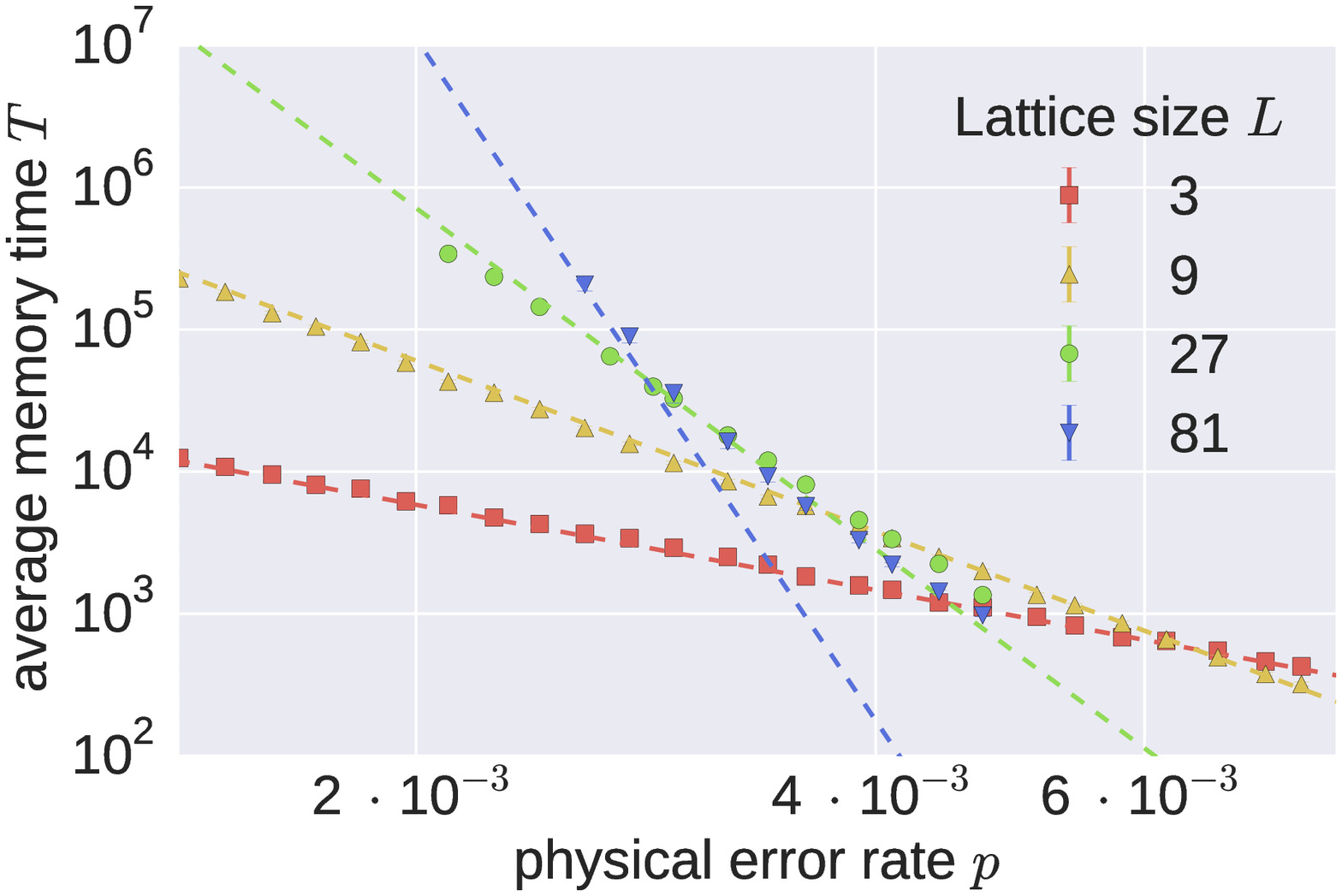} \\
(a) & (b)
\end{tabular}
\fcaption{(Color Online) Average memory time of the decoder at $q=p$ (a) and $q=0$ (b) for different system sizes. The dashed lines are fits given by Eq.~(\ref{eq:ansatz1}).}
\label{fig:memTime}
\end{figure}

Fig.~\ref{fig:memTime} shows data on the average memory time of the decoder for different system sizes $L \in \{3,9,27,81\}$ and single error probabilities $0.07\%< p =q<0.4\%$. Unfortunately the different lines do not cross at a single point (which would more clearly indicate a threshold). However, using ideas from concatenated coding and arguments in Harrington's thesis, we can model the probability for logical failure. In particular, we fit the memory time $T$ depending on error probability $p$ and system size $L$ to the following function:
\begin{equation}\label{eq:ansatz1}
 T = \frac{U^k}{AB^{2^k-2}}p^{-2^k} \ \ ,
\end{equation}
where $k = \log_3(L)$ is the number of hierarchy levels used. This ansatz is motivated by the hierarchical structure of the decoder: Harrington argues this behavior for larger systems ($Q=17$, $U=400$). Roughly, the idea is to assume that a space-time box of $O(1)$ size (related to $Q$ and $U$) corrects a single isolated qubit or measurement error. Thus it requires two such errors to produce a level-1 error indicated by a level-1 syndrome. A concatenated code which can correct a single error then obeys the scaling for the $k$-level error probability $p_k= p_c \left(\frac{p}{p_c}\right)^{2^k}$ where $p_c$ is fixed by the break-even equation at level-1, namely $p_c= B p_c^2$ (see e.g. \cite{AGP:ft}). Here $B$ is a combinatorial factor which counts how many pairs of double errors in the space-time box lead to a level-1 error. The probability $p_k$ would thus be the logical failure probability for a space-time box with work period $U^k$. Thus the logical error probability per unit time is $\frac{p_k}{U^k}$ and $T \sim \frac{U^k}{p_k}$ leading to the ansatz of Eq.~(\ref{eq:ansatz1}). We do not claim that we have a rigorous underpinning of this behavior as the fault-tolerance analysis of this decoder is fairly complex (see also the discussion in Section \ref{sec:physics}).

The dashed lines in Fig.~\ref{fig:memTime} show fits of the data to the above function with $A = 1.0\cdot10^{3}$ and $B = 7.5\cdot10^{2}$ giving rise to a threshold of $p_c = \frac{1}{B} = (0.133\pm0.001)\%$.\\

A similar analysis can be performed when syndrome measurements are perfect $(q=0)$. Again we can fit the data using Eq.~(\ref{eq:ansatz1}) giving rise to parameters $A = 3.9\cdot10^{2}$ and $B = 4.7\cdot10^{2}$ and a threshold of $p_c = \frac{1}{B} = (0.214\pm0.002)\%$.

\begin{figure}[h]
\centering
\begin{tabular}{c c}
\includegraphics[width=0.45\textwidth]{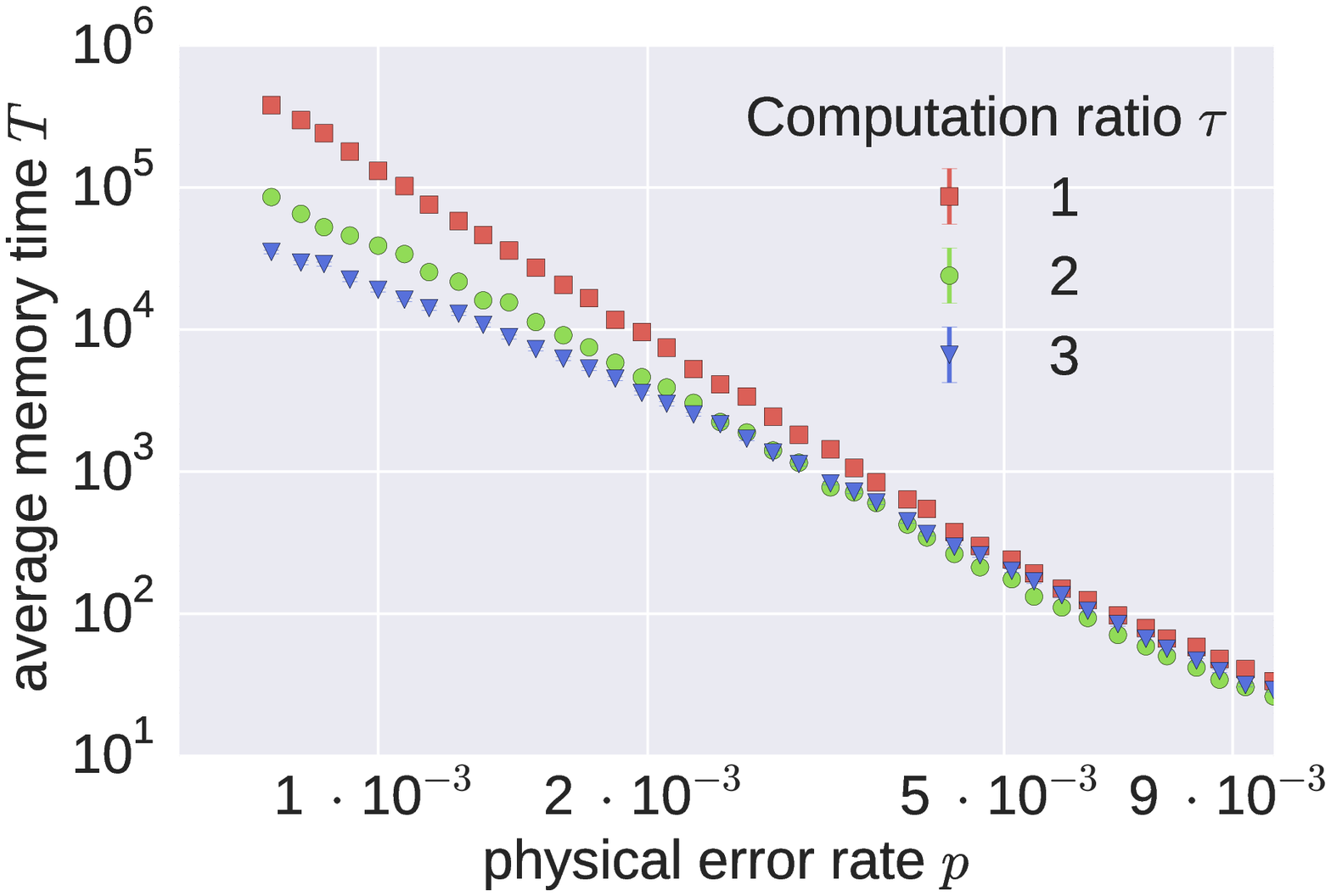} \hspace{1cm} & \includegraphics[width=0.45\textwidth]{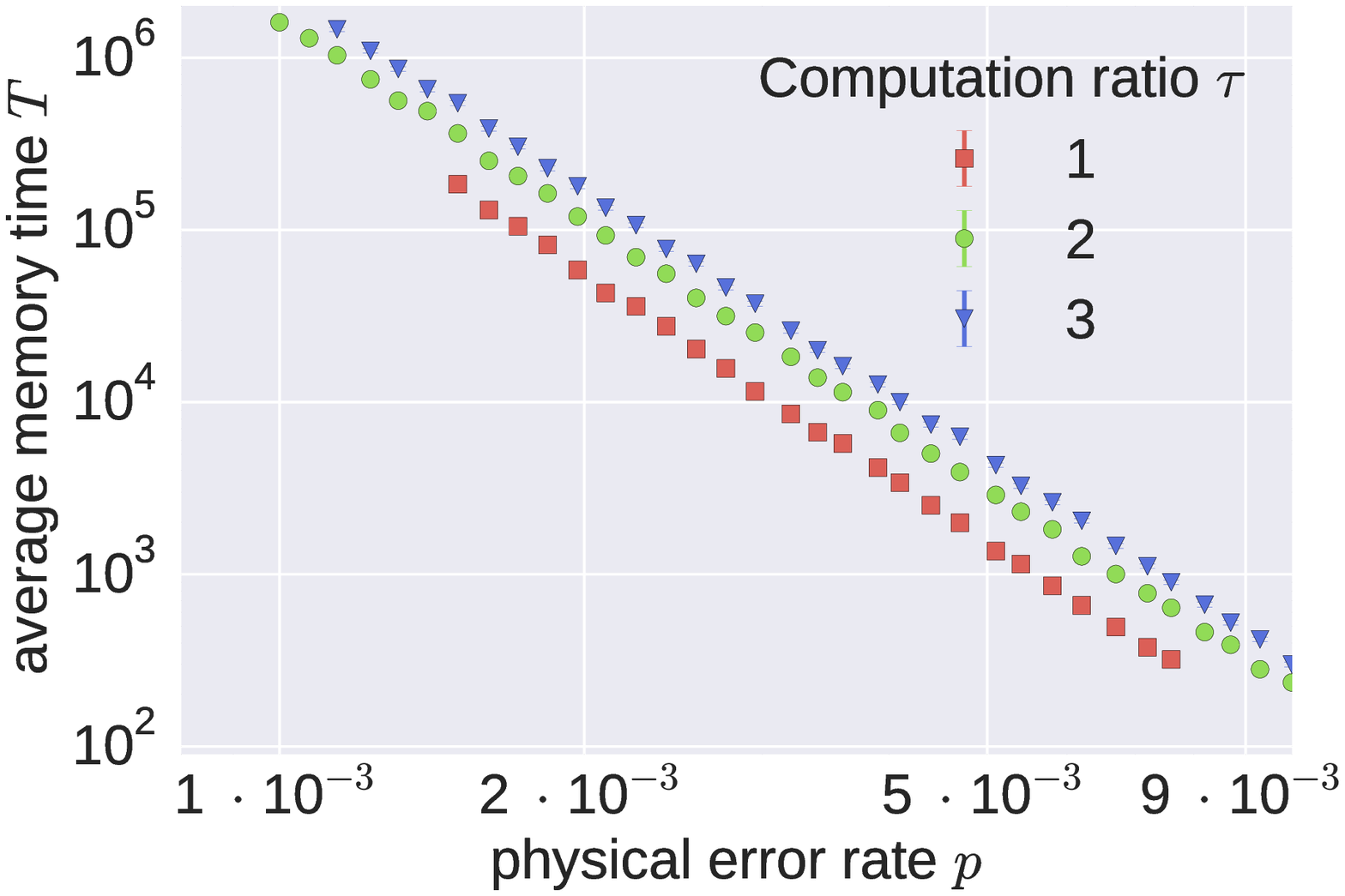}\\
(a) & (b)
\end{tabular}
\fcaption{(Color Online) Average memory time for various computation ratios $\tau$ at $q=p$ (a) and $q=0$ (b). In both cases $Q=3$ and $L=9$.}
\label{im:taus} 
\end{figure}

\subsection{Varying Classical Computation Speed}
\label{sec:speed}

\begin{figure}[htb]
\centering
\includegraphics[width=0.48\textwidth]{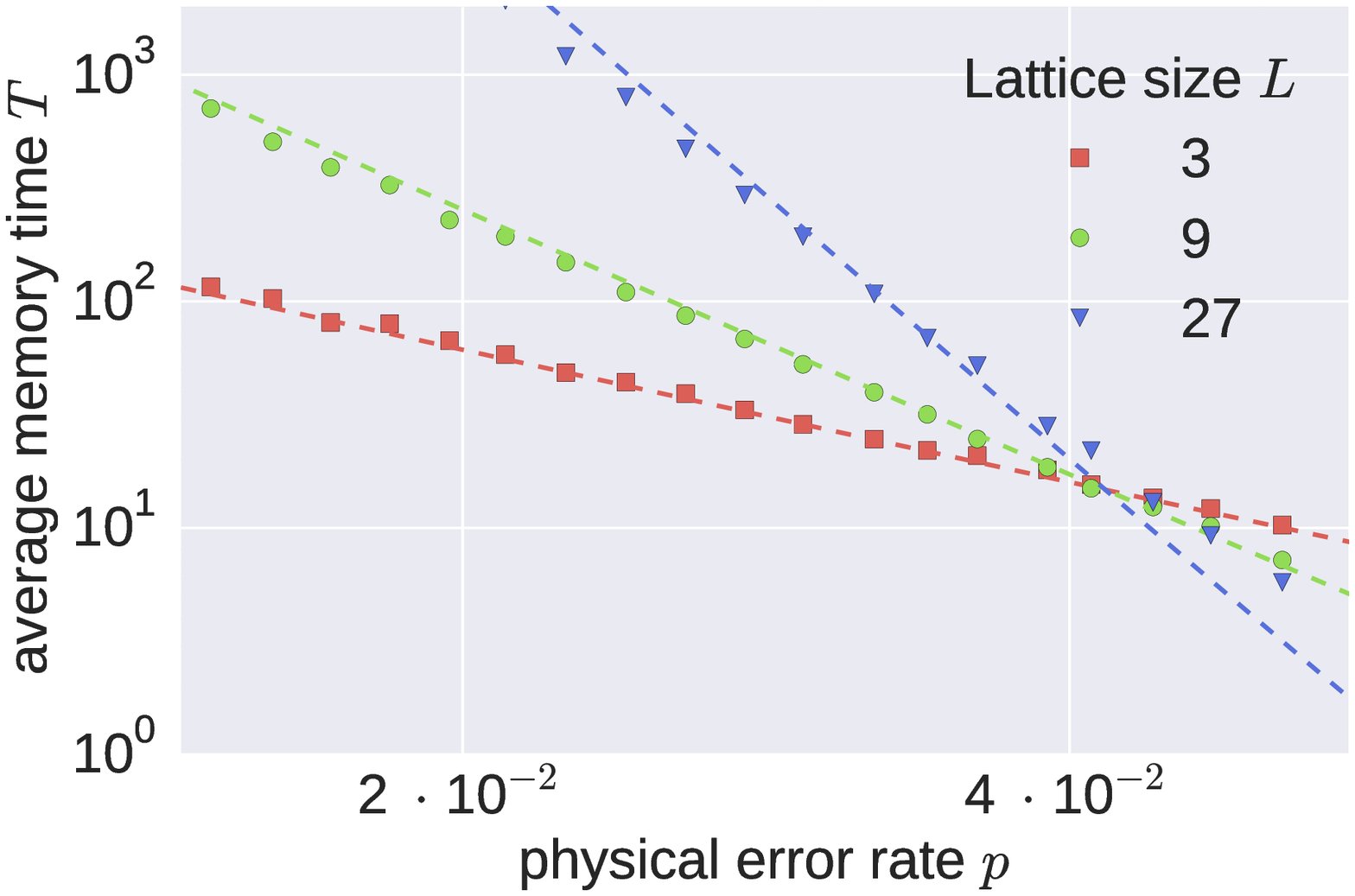}
\fcaption{(Color Online) Average memory time for
$\tau\rightarrow\infty$ at $q=0$. A threshold is
observed around $4.0\%$. The dashed lines are fits given by Eq.~(\ref{eq:ansatz1}) with $U=1$.}
\label{im:decay_tauinf} 
\end{figure}

In the original description of the Harrington decoder, the CA rules and the QEC
cycle work in lock step: every time step the decoder
gets a new syndrome and performs a single run of the local update rules. However, there is no need to tie the speed of 
classical computation to, say, a $5$MHz syndrome acquisition rate. Present-day classical logic has 
at least a $O(1)$GHz clock cycle, so $10^3$ basic logic CA updates for every QEC round may be possible.

We thus assume that the decoder is able to perform $\tau$ correction steps before a new syndrome is coming in and consider
the improvement in performance. We ran the
simulations for different computation ratio $\tau$ and the results are displayed in Fig.~\ref{im:taus}. If the syndrome information is trustworthy, i.e. $q=0$, speeding
up the decoder is quite good because the   decoder is able to correct errors up to a
certain size depending on $\tau$. We simulate the decoder for $q=0$ and $\tau\rightarrow\infty$ (i.e. perform as many decoding steps as needed to remove all syndromes before introducing new errors). We fit the data using Eq.~(\ref{eq:ansatz1}) with $U=1$ since in single time step errors are either all removed or create a logical error. We estimate $A = 43$ and $B = 25$ giving rise to a threshold at $(4.0\pm0.1)\%$ which is significantly higher than the $0.214\%$ for no decoder speedup (see Fig.~\ref{im:decay_tauinf}).

On the negative side, we observe that for $q=p$ speeding up the
decoder decreases the memory time. In this case a single measurement error can propagate over a distance depending
on $\tau$ introducing physical errors. Merely speeding up the decoder in between noisy syndrome acquisition is thus useless as noisy syndromes are removed 
by considering defects in time which requires defects to be moved slowly. In this sense the RG decoder in \cite{DP:3Ddecoding} which considers the syndrome record in time as well as space would be more effective.

\section{Discussion: Error Correction by a Local Physical System}
\label{sec:physics}
In the current form of the Harrington decoder the maximal memory of a 0-cell is $O(\log L)$ since the 0-cell which represents the highest-level $k$-cell should at least contain a clock register which counts from $0$ to the level-k work period, $U^k=U^{\log_Q (L)}$ as well as a register denoting its location in the supercolony of size at most $Q^k$. This implies that the CA decoder does not fully represent a 2D physical system which has $O(1)$ local memory, purely local connectivity and finite speed of communication. Showing that such physical error-correcting system exists has been of some interest as it would establish the quantum counterpart of the Gacs work in 1D. 

Instead of choosing a 0-cell as a physical representative of an $i$-cell, one could instead distribute all the memory and computation steps over the (super) colony that the $i$-cell is receiving its syndrome from.  It may be clear that the functioning of the decoder is not fundamentally altered by letting the logical functioning of, say, the $1$-cell be executed by the underlying $Q^2$ 0-cells, each with $O(1)$ memory. Such modification would simply result in a slower operation of the $1$-cell. The logical functioning of the $2$-cell can then identically be broken down into computation and communication involving 1-cells with $O(1)$ memory, which are again broken down into actions by the $0$-cells. However, to do this using a cellular automaton in which all cells have the same update rules is rather complicated as the Gacs work demonstrates.

A quantum counterpart of the Gacs work could also be established via the use of concatenated quantum error correction: here we discuss this idea. It is known that one can perform fault-tolerant computation in 1D \cite{gottesman:1D}, see also the more explicit construction in 2D in \cite{SDT:local}, assuming that one has arbitrarily fast and noiseless classical decoding. But it is a simple step to include the noiseless decoding procedure as part of the entirely local 2D quantum computation.
Take the concatenated $[[7,1,3]]$ scheme in \cite{SDT:local} and include in each unit cell representing an encoded qubit, the bits and logic required to do classical decoding using the syndrome of the QEC cycle. We thus break down this $O(1)$ amount of classical computation into a $O(1)$ amount of classical local 2D computation. Since the decoding is no longer instantaneous but takes some $O(1)$ time, the idling qubits may undergo additional errors, but this effect can be incorporated in the error model on the qubits. Now in the concatenation step each qubit in the unit cell gets replaced by a unit cell itself. Each logical gate between qubits gets broken down into sequences of local two-qubit gates between qubits in the respective unit cells followed by error correction in the unit cell. Since the classical bits and gates are assumed to be noiseless, each classical bit can be replaced by a unit cell of the same size as the unit cell representing a qubit, but otherwise consisting simply of noiseless finite-speed communication channels which allow the bit to undergo (two-bit) logical gates with neighboring bits in unit cells. Since each logical gate in the decoding computation now takes longer (due to blow-up in size and finite communication speed), one can assume that a single quantum error correction step is applied on the waiting qubits during this gate. Applying the standard tools of concatenated coding \cite{AGP:ft} to this construction one should get an error probability for the logical qubits which goes down doubly-exponentially with concatenation level assuming one is below a threshold error probability on every physical qubit. If the classical decoder is noisy, these arguments do not directly apply and one should similarly concatenate the classical parts of the computation. The idea would then be to ensure that a single error in either the classical or quantum part of the operations in the unit cell cannot generate 2 (and thus incorrectable) errors on the 7 qubits, following the prescription of fault-tolerance. Thus one expects that formal arguments could also be applied to this case, leading to a lower noise threshold due to the spatial overhead and time-delay incurred by making classical computation robust. 

A technical difficulty in applying the techniques of concatenation coding directly to the setting of 2D topological coding is that the toric code is not a concatenated code, but rather a concatenated code in which boundary logical qubits encoded in patches are removed (renormalized away) in the concatenation step where patches of the lattice (corresponding to colonies) are glued together. It is an open and interesting question whether one can develop techniques such as the ones in \cite{AGP:ft} to analyze the threshold by fully local RG decoding (instead of using error cluster expansions as in \cite{thesis:harrington} and \cite{BH_RGdecoder}).

\section{Decoding the 4D Toric Code}
\label{sec:4Ddecoding}

\subsection{Hastings Decoder}
\label{sec:hastings}

In \cite{hastings} Hastings proposed a local decoder for 4D homological codes based on a hyperbolic tiling \cite{guth_lubotzky}. Using the fact that in negatively curved spaces the size of the boundary of a surface scales proportional to its interior, he proved that 
with the application of the decoder on some $O(1)$ ball (or neighborhood) in the lattice the weight of an error ($i.e.$ the number of qubits on which it acts nontrivially) is reduced by a constant fraction on average. The decoder is thus effective at removing errors in this geometry. We will consider its action 
here on a code obtained from tiling a 4D flat space, namely the 4D toric code.

\begin{figure}[ht]
\centering
\begin{tabular}{c c}
\hspace{1cm}
\includegraphics[width=0.2\textwidth]{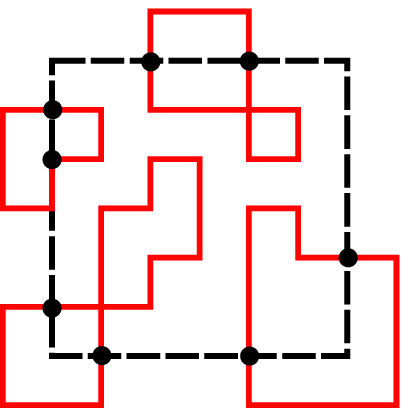} \hspace{1cm} &  \includegraphics[width=0.2\textwidth]{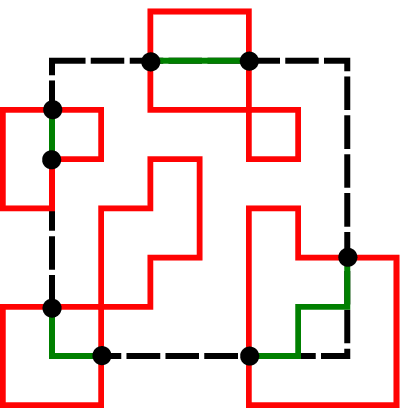}\\
(a) & (b)\\
 & \\
 \hspace{1cm}
\includegraphics[width=0.2\textwidth]{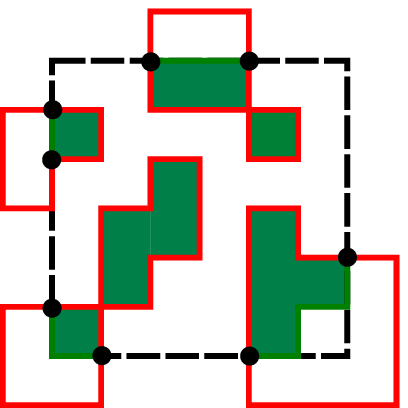} \hspace{1cm} &  \includegraphics[width=0.2\textwidth]{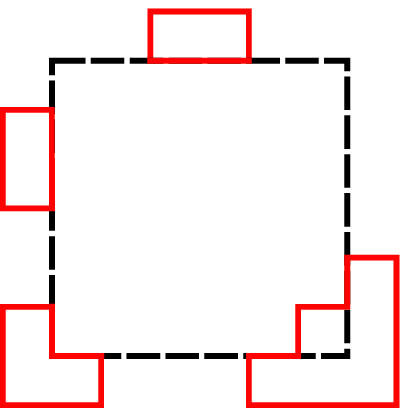}\\
(c) & (d)\\
\end{tabular}
\fcaption{(Color Online) Schematic picture of local error correction procedure for a box with $O(1)$ qubits. (a) Syndrome $S$ (red) and a black box $N$. (b) Find the set of strings $S'$ (green) of minimal length which connects the intersection vertices $V$ (vertices where $S$ intersects the boundary of $N$.) (c) Find a collection of sheets $R$ with minimal area which has the closed loops $S|_{N}+S'$ in the interior of the neighborhood as its boundary. (d) Residual syndrome after the local decoding step.}
\label{fig:local_decoder}
\end{figure}

We first describe the local decoding procedure assuming a perfect measurement of the check operators. In Section \ref{sec:noisy} we show that the same method works when the parity check measurements are subject to noise.

\begin{figure}[ht]
\centering
\includegraphics[scale=0.35]{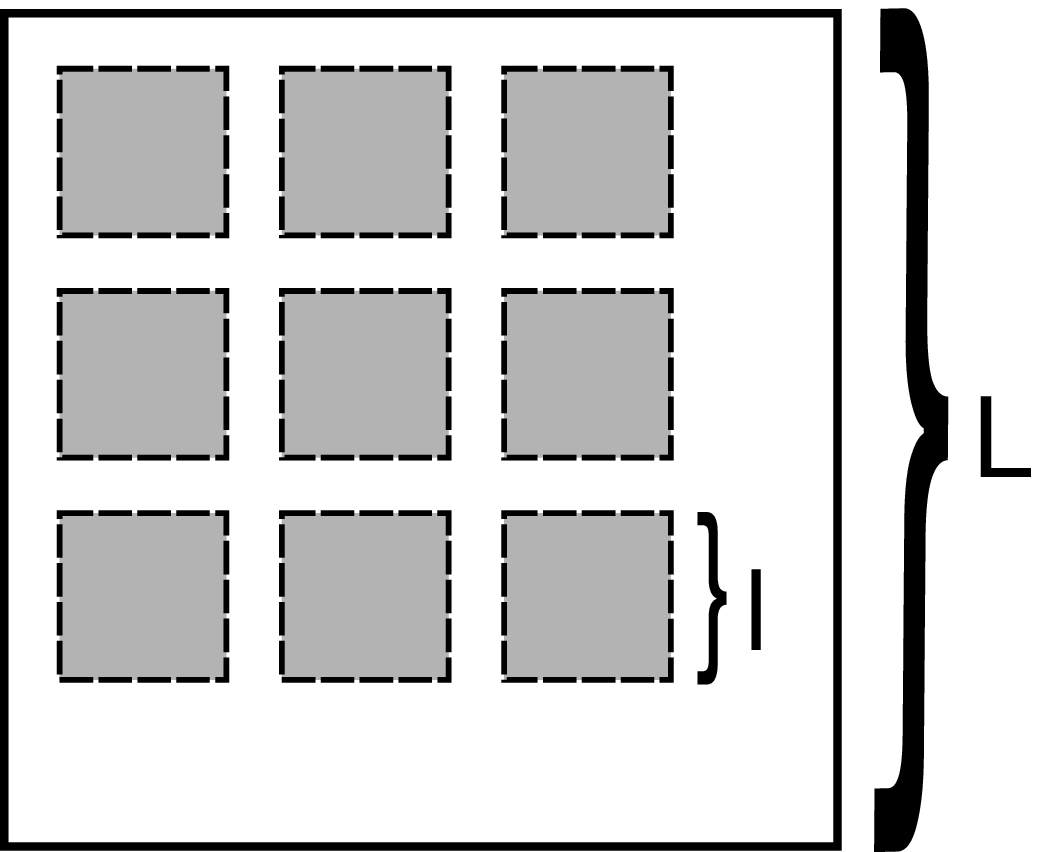}
\fcaption{(Color Online) Layout of the neighborhoods in the 4D hypercubic lattice of side-length $L$. Each neighborhood is a box with side-length $l$. In each decoding round the location of the grid of boxes is chosen at random.}
\label{fig:boxlayout}
\end{figure}

Imagine that the lattice is split up in non-overlapping hypercubic boxes, each box $N$ defining a subset of the lattice (i.e. a collection of vertices, edges, faces, cubes and hypercubes which form a connected region of the lattice). For the hypercubic lattice, the optimal choice is to take each box of side-length $l$ and placing the boxes in a grid (see Fig. \ref{fig:boxlayout}). This means that we can fit
\begin{equation}\label{eqn:numboxes}
\lfloor L/(l+1) \rfloor^{4}
\end{equation}
boxes of side-length $l$ into a hypercubic lattice on a torus of size $L^{4}$.  Note that at the boundary of the box some of the qubits on which an edge check operator acts do not need to be included inside the box, and check operators from different boxes may act on the same qubits (both of which are outside the box).

Since this arrangement does not cover the whole lattice (i.e. include all qubits) we will change this partition of the lattice a few times in the decoding process. Consider a single box $N$ of fixed $O(1)$ size, not scaling with $L$.
Denote the set of edge check operators which have non-trivial syndrome and which are contained in $N$ by $S|_{N}$. $S|_N$ may consist of closed loops which are completely confined within $N$ and for example open strings which pass through the boundary of $N$, see Fig.~\ref{fig:local_decoder}(a). 
In particular given $S|_N$ we define the set of (intersection) vertices $V \in N$ as vertices in $N$ which are the boundary to an {\em odd} number of elements in $S|_N$. $|V|$ is always even.

The first step of the decoder is to determine the shortest distance matching (MWM) between the vertices $V$, see Fig.~\ref{fig:local_decoder}(b). 
The matching is done using the Euclidean distance $\mbox{dist}(\vec{x},\vec{y})=\sqrt{\sum_{i=1}^4 (x_i-y_i)^2}$. We choose the path of edges connecting pairs of matched vertices to be the one which deviates least from the direct path. If this should be the case for more than one path of edges, we pick one uniform at random among this set. This step will always keep the length of the non-trivial syndrome in $N$ the same or shorten it.

\begin{figure}[ht]
\centering
\begin{tabular}{c c}
\hspace{0.5cm} \includegraphics[width=0.2\textwidth]{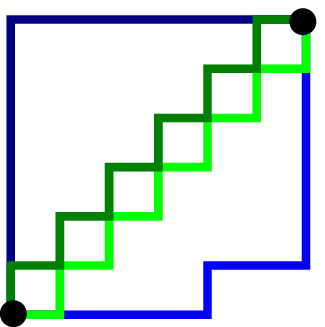} \hspace{0.5cm} & \hspace{0.5cm} \includegraphics[width=0.2\textwidth]{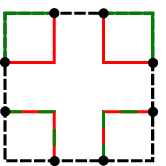}\hspace{0.5cm}\\
(a)  &  (b) 
\end{tabular}
\fcaption{(Color Online) Paths between pairs of matched vertices. (a) According to the taxi-cab metric all four lines connecting the two dots have the same length equal to 12. But among all paths of length 12, the green ones approximate the direct path best. (b) Syndrome (red) entering box with side length $l=3$ at four corners. The shortest distance matching is either along the boundary (upper corners) or through the interior (lower corners). Hence only on the faces in the upper corners a correction is applied.}
\label{fig:eucl_matching}
\end{figure}

Note that using the Euclidean distance is different from taking a taxi-cab norm on the lattice or graph where one just counts the path length in terms of edges between a pair of vertices.  Choosing the taxi-cab distance would not be effective in shrinking error regions as depicted in Fig.~\ref{fig:eucl_matching}(a). For small syndrome loops at the boundary, see Fig.~\ref{fig:eucl_matching}(b), taking a random path of minimal length $d$ will act similarly as the DKLP rule for a qubit face surrounded by two non-trivial checks.

Let $S'$ be the resulting set of edges and note that $S_{\rm correct}=S'+S|_N$ has the property that all vertices in $N$ touch an even number of elements of $S_{\rm correct}$, hence $S_{\rm correct}$ consists of only closed loops. Note that where $S'$ and $S|_N$ coincide they will cancel and they enclose no area. 

In the next step we determine the smallest surface $R$, i.e. a subset of faces in $N$, which has $S_{\rm correct}$ as its boundary, see Fig. \ref{fig:local_decoder}(c) and is found by solving an integer program (integer programming is not computationally-efficient but the box is of $O(1)$  size and integer programming is preferred over brute force enumeration.) $R$ is thus the proposed $Z$ correction for the box. If we would $Z$-flip all qubits corresponding to $R$, we are left with a residual syndrome as in Fig. \ref{fig:local_decoder}(d).

The decoder applies this procedure in parallel on each box and each procedure in a box clearly takes $O(1)$ computation time, hence the decoder is local. The parallel action results in a total correction $R_{\rm total}$. Before applying $R_{\rm total}$ or recording it as the final correction, we can repeat this decoding procedure with a different box-partition. This is useful since a partition leaves some qubits outside every box, hence no correction can take place on them. 

Thus we allow the decoder to re-apply the procedure $m$ rounds, with each round being a different partition (but keeping boxes of the same size). Implicitly, it means that we allow the classical decoder to become have high computational speed. After every round the total syndrome is updated given the current recovery and the next round is applied to the left-over syndrome. 

\subsection{Noisy Syndrome}
\label{sec:noisy}

\begin{figure}[ht]
\centering
\begin{tabular}{c c}
\hspace{0.5cm}
\includegraphics[width=0.2\textwidth]{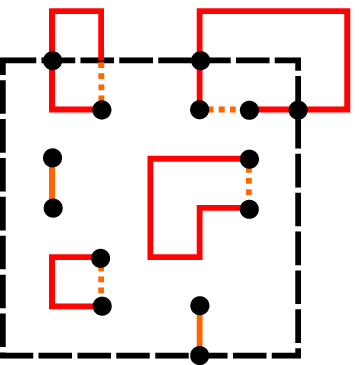} \hspace{0.5cm} & \hspace{0.5cm} \includegraphics[width=0.2\textwidth]{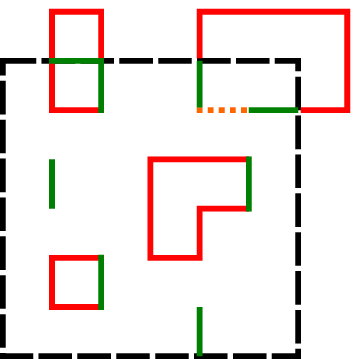} \hspace{0.5cm} \\
(a)  &  (b) 
\end{tabular}
\fcaption{(Color Online) Schematic picture of local error correction procedure in the presence of erroneous syndrome measurements. (a) Local decoding neighborhood in the presence of error syndromes (red) and syndrome errors (orange). The measured syndrome consists of solid lines while dashed lines indicate edges where the syndrome error overlaps with the real syndrome. (b) After performing a matching (green lines) we are left with a collection of closed loops. We are in the same situation as in Fig.~\ref{fig:local_decoder}(b) and can continue with the correction. Note that edges where the measured syndrome and the matching overlap are removed.}
\label{fig:locel_dec_noisy}
\end{figure}

When the syndrome is noisy, the measured syndrome is a collection of open strings instead of closed loops. It is possible to determine the 0-dimensional boundary of this collection and perform global MWM on this set of boundary vertices, but this would result in a {\em non-local} single-shot decoder. Instead we can apply the decoding procedure described above: the vertex set $V$ now simply includes vertices in the interior of $N$, see Fig.~\ref{fig:locel_dec_noisy}(a) and Fig.~\ref{fig:locel_dec_noisy}(b).  One is left with a collection of closed loops as in Fig. \ref{fig:locel_dec_noisy}(c) for which we can again proceed as before. 

The decoding procedure is single-shot in the sense that we only use the data obtained from a single round of syndrome measurements and the redundancy in the syndrome is used to repair the syndrome record. 
We note that doing the decoding in two steps, namely first syndrome repair and then error inference, can be done for the 3D gauge color code as well due to the redundancy in the gauge check syndrome record: in \cite{BNB:singleshot} a (non-local) clustering decoder was used to repair the syndrome record.

\subsection{4D Toom and DKLP Decoders}
\label{sec:toom_dklp}

We first discuss Toom's rule which was introduced in \cite{toom}. Consider a 2D square lattice where each face is associated with a degree of freedom which can take the values $+1$ and $-1$. Each face is surrounded by four edges which we will label `north', `east', `south' and `west'. Two neighboring faces share exactly one edge which is either in the east of one face and the west of the other or in the north of one and in the south of the other. Every edge is associated with the parity between the two faces incident to it. An edge-check is non-trivial if and only if the two faces incident have different values. Toom's rule states that for each face the value is flipped if and only if the parity checks of the north and the east edges are non-trivial. If we view Toom's rule as a cellular automaton, it means that the cell of the automaton resides on faces, above the qubit, and the automaton processes the syndrome of its NE edges (compare with the cells of the Harrington decoder).
 
 To turn Toom's rule into a decoder for the 4D toric code we can for example apply the update rule to every 2D plane of the 4D hypercubic lattice. We thus partition the set of all planes into 6 groups, each of a set of parallel planes. We then apply the rule on the first group, say, on all $x$-$y$-planes, then the second group, i.e. all $x$-$z$-planes, etc. Within each group the CA rule is applied on all qubit faces in parallel. One needs to fix an orientation in the lattice so that the vertex between the `North' and `East' edges is the vertex with the largest lattice coordinates (modulo L). After this round of applications a new syndrome record can be obtained. It may also be possible to apply Toom's rule in parallel on all qubits in the lattice, but we have not implemented this.

The DKLP rule counts the number of non-trivial edge-checks and does a majority vote as described in the Introduction. The idea behind the rule is that the weight of the non-trivial syndrome is never increased but it may be decreased. According to \cite{DKLP} the update rule can be applied to a set of faces which do not share a common edge. Thus we partition the lattice in these 6 planar groups as for Toom's rule, but then we further subdivide each plane into sets of non-overlapping faces by dividing them into a checkerboard pattern and apply the rule only in parallel on each non-overlapping set.

\subsection{Energy-Barrier Limited Decoding}
\label{sec:barrier}

\begin{figure}[h]
\centering
\includegraphics[scale=0.35]{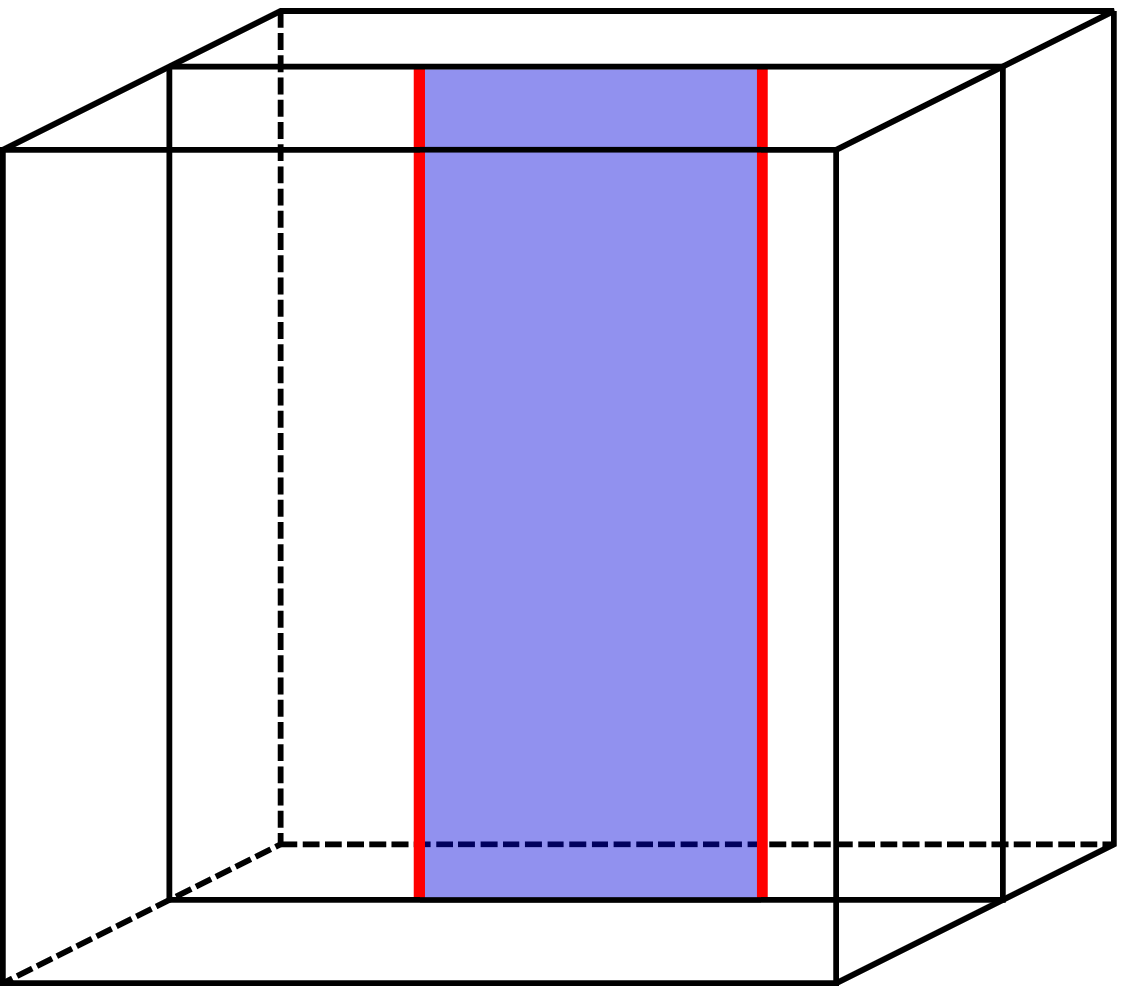}
\fcaption{(Color Online) A 2D slice in a 3D slice in the full hypercubic lattice where opposite sides are identified. A sheet of errors (blue) gives rise to a syndrome (red) on which the 4D Toom, DKLP and Hasting decoders can get stuck.}
\label{fig:stuck_error}
\end{figure}

For all decoders discussed in the previous sections, it can be observed that the decoders are unable to shrink and remove certain high-weight syndromes (assuming for simplicity that syndromes are noiseless). This implies that these decoders cannot necessarily correct a state with errors back to a codeword. An example of such  syndrome can be seen in Fig.~\ref{fig:stuck_error} where we imagine looking at a 2D plane of qubit faces. The qubit faces are flipped along a homologically non-trivial strip. For the Toom's rule decoder it can be observed that a full column of flipped qubits will not change as all qubits have even parity with the north qubit and odd parity with the east qubit. If the flipped qubits form a single column, the DKLP decoder will still apply corrections at random since every qubit in the column has two non-trivial edge checks.  Similarly, the Hastings decoder implements flips if the box size is sufficiently large ($l\geq 2$) so that intersection vertices at the top get matched together and separately intersection vertices at the bottom. However, if one makes the strip of qubit errors two-columns wide, the DKLP decoder will no longer apply corrections as every qubit sees only one non-trivial check. Similarly, when the strip width $w \geq l$ where $l$ is the side-length in the Hastings decoder, then the string intersecting the boxes is locally of minimal length. Making the box larger could thus be useful, but comes at increased non-locality and computational resources. A non-local decoder which finds the minimal surface matching the closed loop boundary would be able to correct these errors.

In such cases when the decoder get stuck and the syndrome does not change, we call logical failure. In Section \ref{sec:numerics4D} we show that for the Hastings decoder this mode of failure is very common, becoming more dominant with increased lattice size. Note that this failure mode is related to the energy barrier \cite{BT:mem} for the code: the errors that produce the nonlocal syndrome that the decoder cannot handle are precisely the errors which set the height of the energy barrier, namely $2L$ for the 4D toric code. The strip of errors can grow, without anti-commuting with yet more edge check operators, to become a logical operator which covers the whole 2D surface. Hence, once the error probability is high enough that errors are generated which locally have minimal energy (anti-commute with the minimal number of check operators), the decoder starts to fail. We don't know whether these error strips are currently an important failure mode of the Toom's rule and DKLP decoders.

\section{Numerical Results With The 4D Decoders}
\label{sec:numerics4D}

\subsection{Numerical results for the Hastings decoder}

In our simulation we choose the boxes of side length $l=3$ which is the smallest non-trivial size (for $l=2$ there is not much shrinking of syndrome loops that the decoder can do). The number of faces of a box of side-length $l$ equals $6l^4 + 12l^3 + 6l^2$. Hence, for $l=3$ each box includes 864 qubits. This shows that the minimal box version of the Hastings decoder is substantially less local than the Toom or DKLP decoders.

The lattice sizes which are numerically tractable and support more than one box at a time are the lattices with $L\in \lbrace 8,9,10,11 \rbrace$ which by Eq.~(\ref{eqn:numboxes}) all support 16 boxes. During a single time step/QEC cycle the position of the grid of the boxes (as depicted in Fig.~\ref{fig:boxlayout}) is varied randomly five times and each time the decoding process is run inside each neighborhood. We chose the number of rounds to be $m=5$ as any increase did not improve the decoder's performance. Unfortunately, due to limited computational resources, we were not able to test the performance of the decoder on larger lattices with an increasing number of boxes. This would have been useful to validate that the number of rounds per QEC cycle can be taken to be independent of $L$ indeed.

To check whether we can restore the encoded data we check after each QEC cycle plus round of corrections whether we can correct back to a code word (assuming a perfect syndrome record). Ideally, this step would be implemented by a nonlocal minimum-weight decoder which finds a minimal surface, but brute forcing this is computationally infeasible. Thus we replace this by letting the local decoding procedure run indefinitely with perfect syndrome measurement.  If we can restore back to a code word without having applied a logical operator we call no logical failure and continue with the next time step. Otherwise we call failure and save the number of time steps until failure. In the end we obtain a value for the average number of time steps until the decoder fails depending on the physical error rate $p$ and the lattice size $L$.

To determine the value of the critical physical error $p_c$ we assume a scaling behavior in the variable $x=(p-p_c)L^{1/\nu}$. For $p$ close to $p_c$ the memory time is well approximated by a quadratic polynomial in the scaling variable. Hence, we do a global fit of the function
\begin{equation}\label{eqn:4Dquadfit}
T(p,L) = T_c + Ax + Bx^2 
\end{equation}
to our data with fitting parameters $T_c$, $p_c$, $\nu$, $A$ and $B$, similar as it has been done for the 2D toric code in \cite{WHP:threshold}.

\begin{figure}[ht]
\centering
\begin{tabular}{c c}
\hspace{0.5cm}
\includegraphics[width=0.45\textwidth]{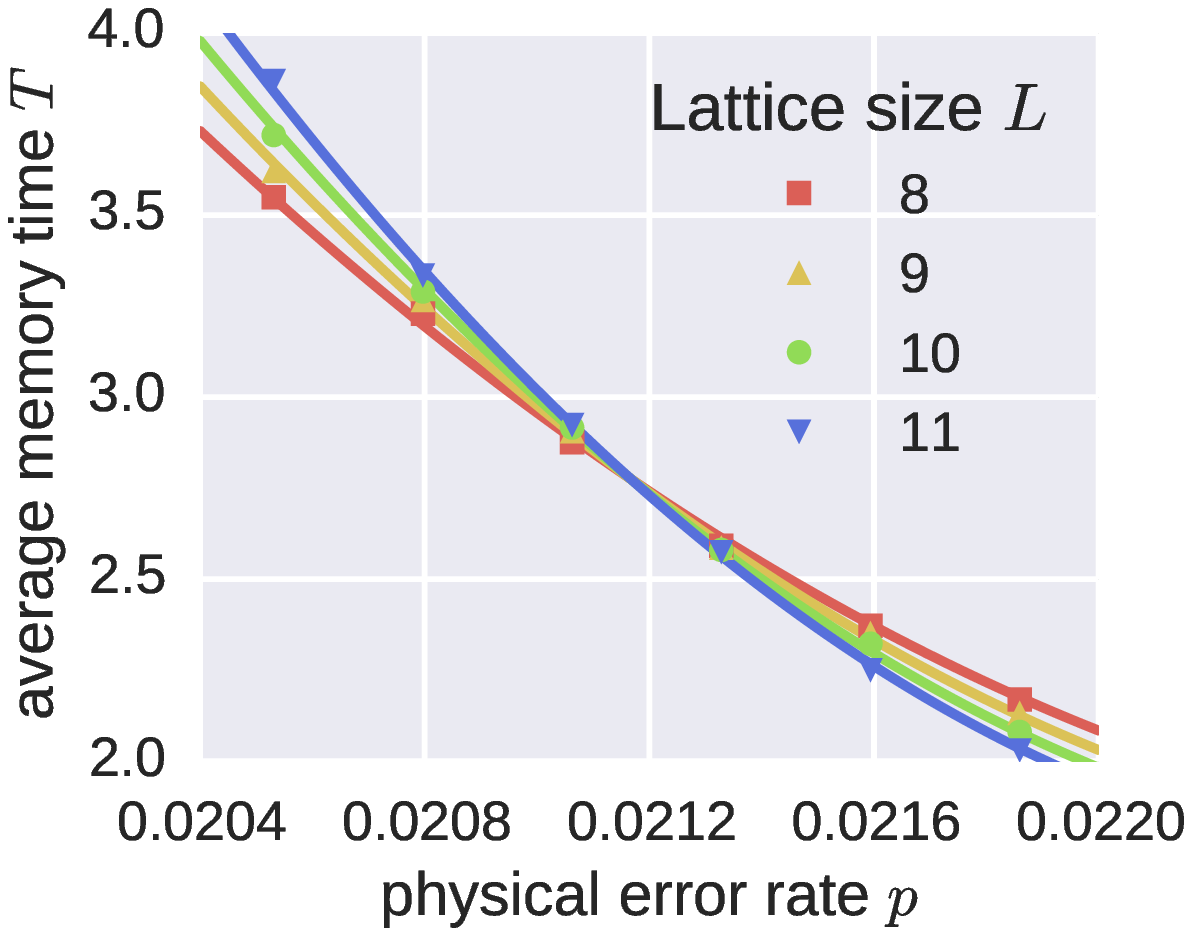} \hspace{0.25cm} & \hspace{0.25cm} \includegraphics[width=0.43\textwidth]{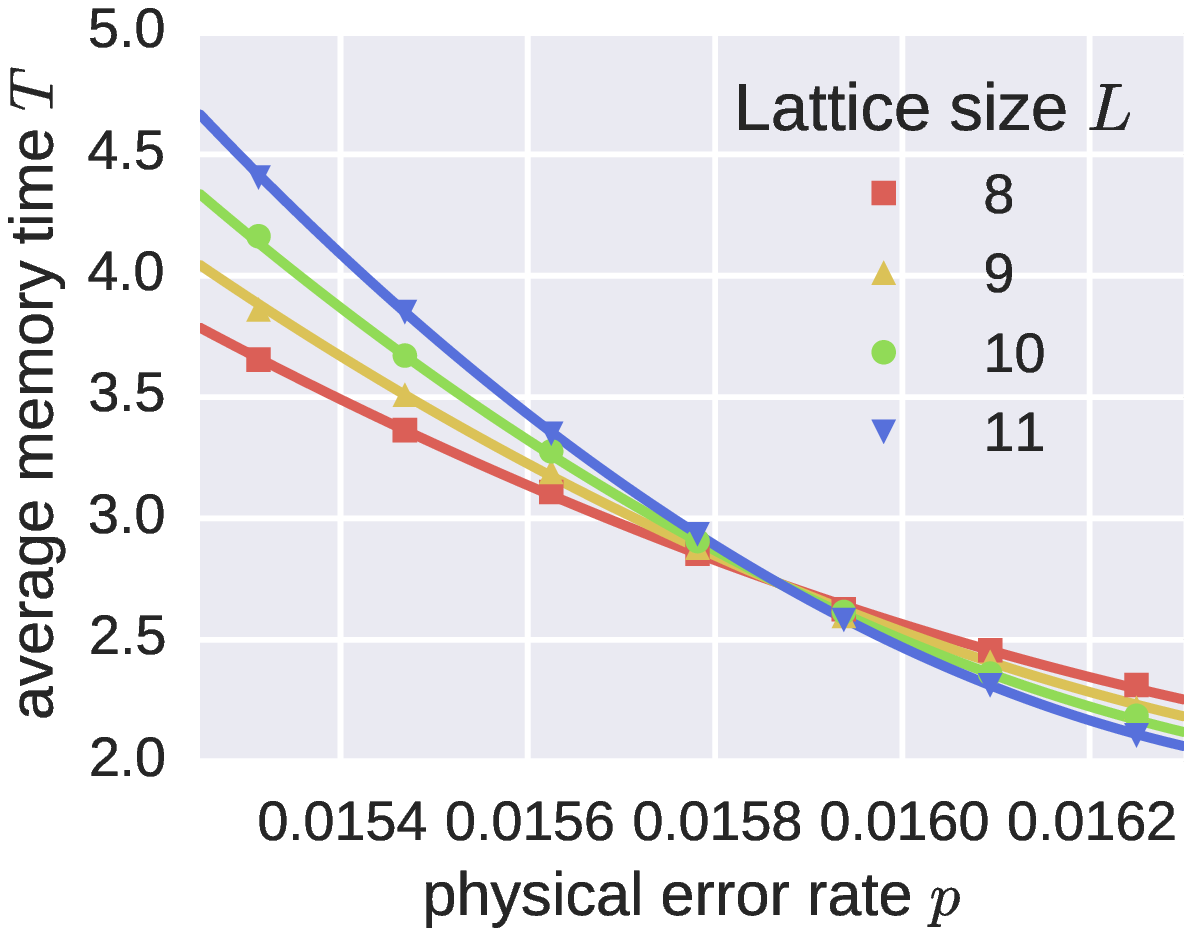} \hspace{0.5cm} \\
(a)  &  (b) 
\end{tabular}
\fcaption{(Color Online) Memory time $T$ depending on the physical error rate $p$ for the Hastings decoder with (a) perfect syndrome measurement and (b) noisy syndrome measurement. The fit was obtained by via Eq.~(\ref{eqn:4Dquadfit}).}
\label{fig:hastings_results}
\end{figure}

For perfect syndrome measurements we find 
\begin{equation}
p_c = 2.117\% \pm 0.006\%, \\
\nu = 1.14 \pm 0.18.
\end{equation}
The data and the fitted function are shown in Fig.~\ref{fig:hastings_results}(a). The data was obtained by running the simulation $5\times 10^4$ times for each data point.

For the simulations where we do model syndrome errors with probability $q=p$ we find
\begin{equation}
p_c = 1.587\% \pm 0.002 \% , \\
\nu = 0.65 \pm 0.03 .
\end{equation}

The data and fitted function are plotted in Fig.~\ref{fig:hastings_results}(b). The data was obtained by running the simulation $5\times 10^4$ times for each data point.

The value of the threshold only decreases by a factor of roughly $1.3$ as opposed to the 2D toric code where it decreases from around $11\%$ to about $3\%$ \cite{WHP:threshold, fowler2009high, thesis:harrington} which is more than a factor of 3.

\begin{figure}[h]
\centering
\includegraphics[width=0.5\textwidth]{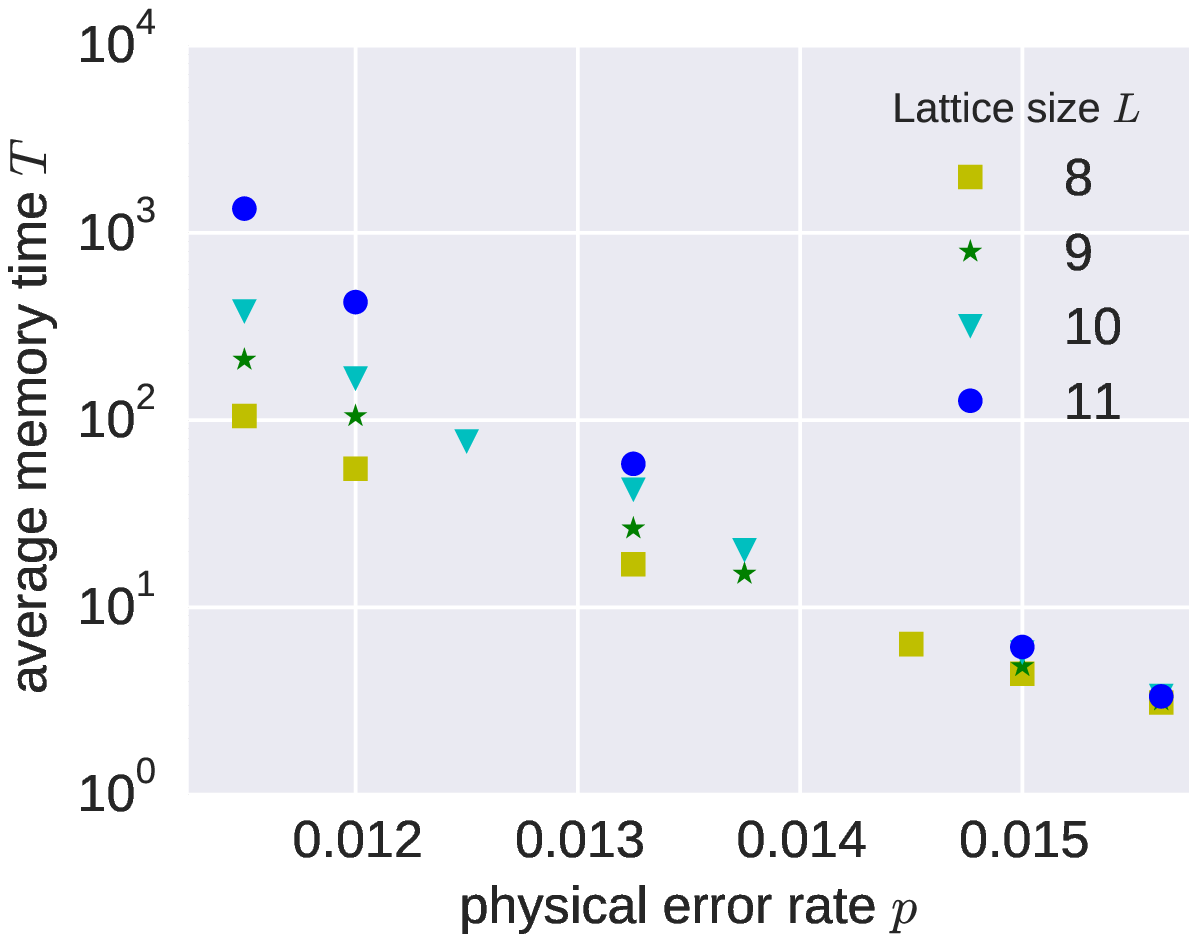}
\fcaption{Some results for the Hastings decoder for smaller values of $p=q$. The average memory time increases up to around $T=1500$ for the $L=11$ lattice for $p=1.15\%$. Due to increasing computational demands we were only able to run around 100 trials for each data point.}
\label{fig:hastings_low_p} 
\end{figure}

Note that the memory times in Fig.~\ref{fig:hastings_results} are extremely small since we are looking at data which are close to threshold. In order to see the trend for much lower error rates, one can look at Fig.~\ref{fig:hastings_low_p} where the number of Monte Carlo trials is relatively low. 

\begin{figure}[htb]
\centering
\includegraphics[width=0.45\textwidth]{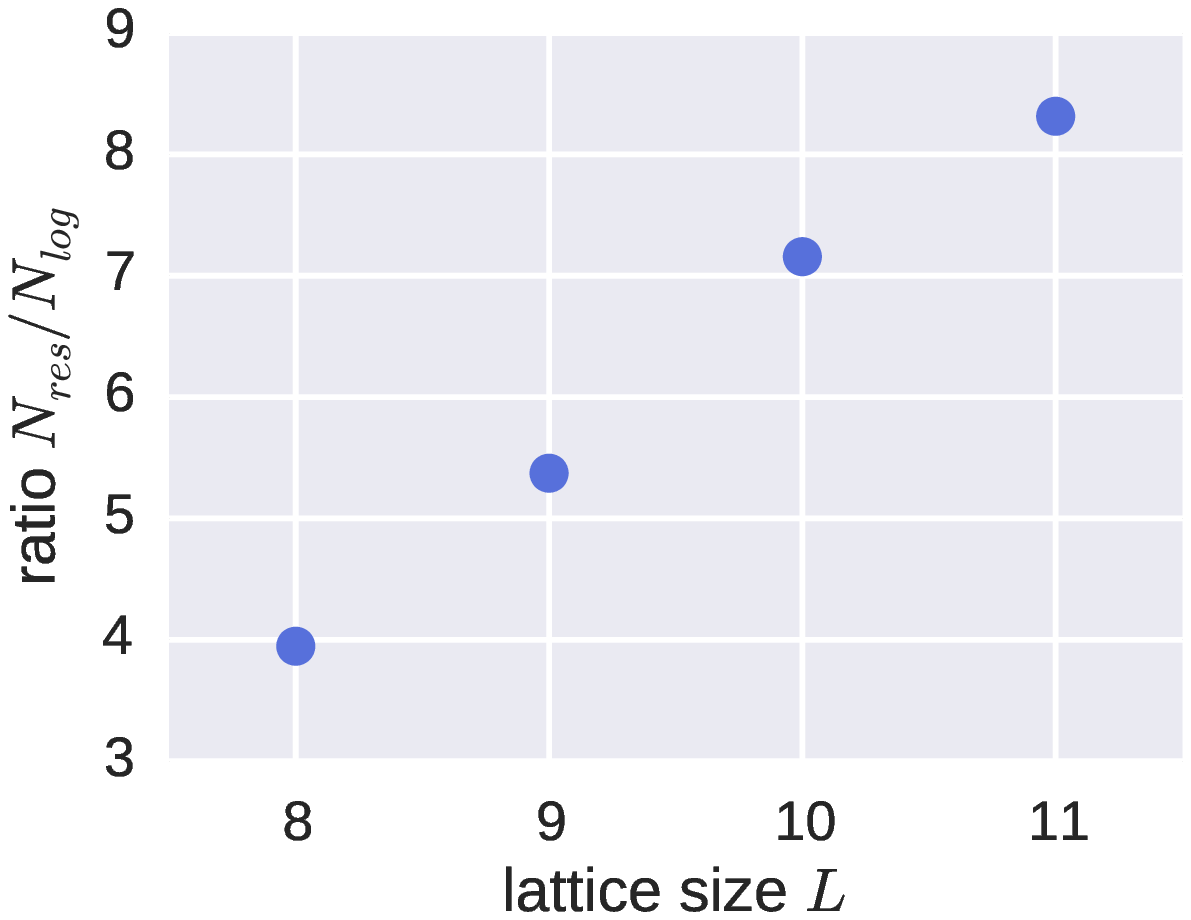}
\fcaption{(Color Online) Dependence of the ratio $N_{res}/N_{log}$ on lattice size $L$ at $p=2.1062\%$ assuming perfect syndrome measurement.}
\label{fig:stuckplot}
\end{figure}

We observe that the number $N_{res}$ of failures of the decoder due to residual syndromes, see Section \ref{sec:barrier}, is higher than the number $N_{log}$ of failures where the decoder did manage to correct back into a code state but applied a logical operator. This may be explained due to the fact that errors for which the local decoder gets stuck require an error of size $O(L)$ to occur in contrast to errors which lead to a logical error which require at least $O(L^2)$ errors, showing that the decoder is energy-barrier limited. This could explain the fact that the ratio $N_{res}/N_{log}$ increases with system size $L$ as can be seen in Fig.~\ref{fig:stuckplot}.

\subsection{Numerical results for the DKLP and Toom's rule decoders}

To simulate the performance of the cellular automaton decoders described in Section \ref{sec:toom_dklp}, we proceed similar as for the Hastings decoder. After applying the decoder we check whether we could correct back to the original code state by running the same decoder indefinitely under perfect syndrome measurements. If no logical operator was applied and the decoder did not get stuck we move on to the next time step.

\begin{figure}[h]
\centering
\begin{tabular}{c c}
 \includegraphics[width=0.45\textwidth]{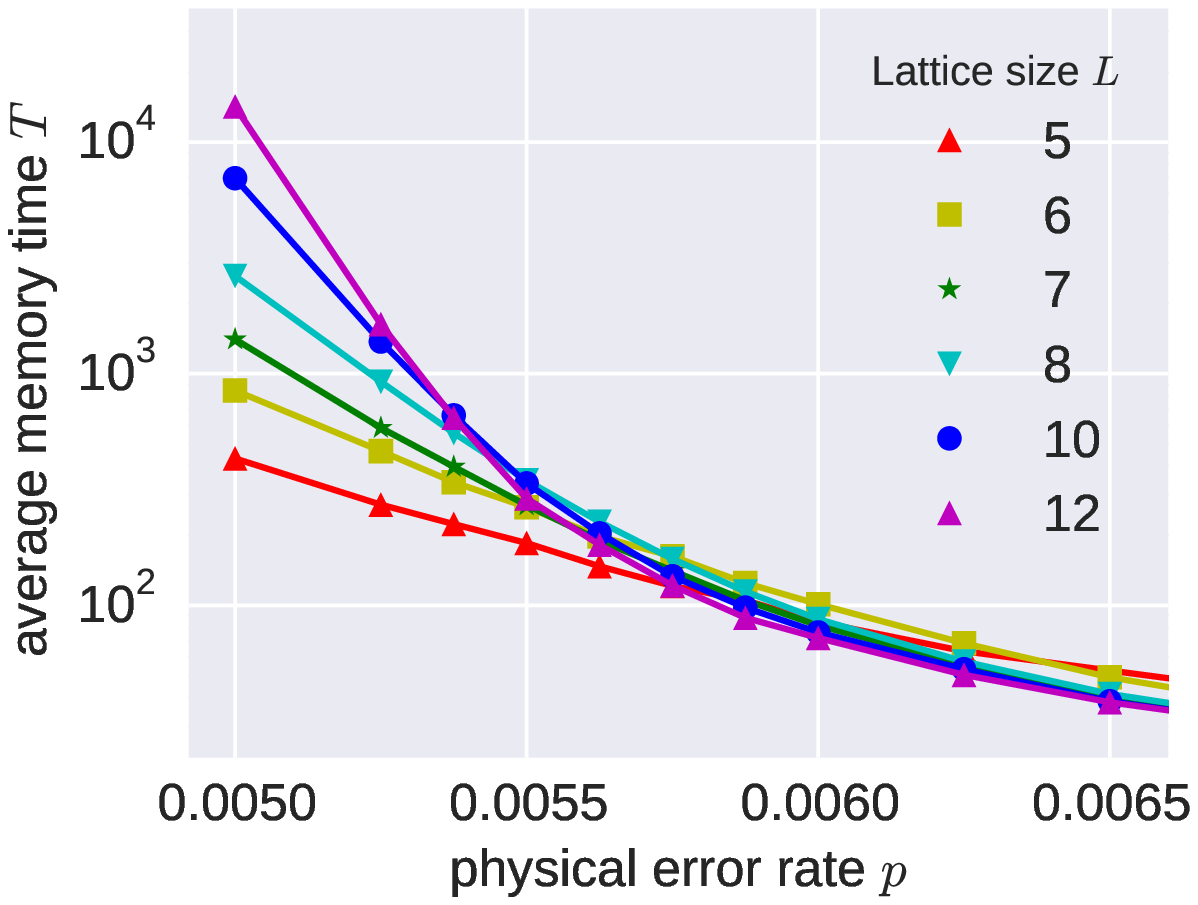}  &  \includegraphics[width=0.45\textwidth]{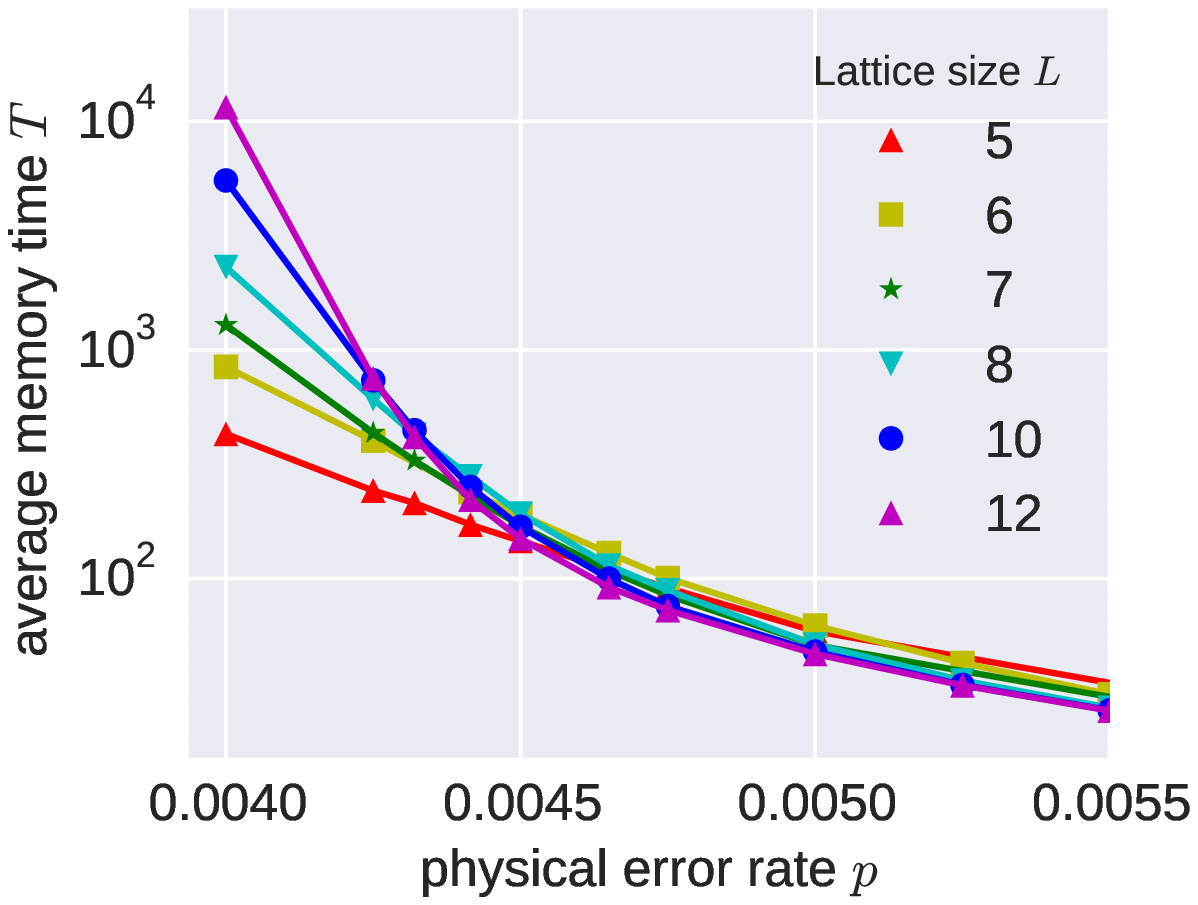} \\
(a) &  (b)
\end{tabular}
\fcaption{(Color Online) Results for the DKLP decoder.  At each time-step the update rule is applied to every plane once. (a) Without syndrome errors  (b) With syndrome errors }
\label{fig:DKLP_results} 
\end{figure}

The results for the DKLP decoder for different lattice sizes can be seen in Fig.~\ref{fig:DKLP_results}. Every data point was obtained by $4\times 10^3$ trials. We observe that for increasing lattice size the crossing point is receding in the direction of lower physical error rate. A finite-size scaling analysis was not possible, as we can only consider relatively small system sizes.

In contrast to the results on the Hastings decoder we see that the average memory time is much higher when we are close to the cross-over point ($O(10^2)$ compared to $O(10^0)$ for the Hastings decoder). For the lattices that we consider the cross-over point occurs between $0.5\%$ and $0.6\%$ without syndrome errors and between $0.4\%$ and $0.5\%$ with syndrome errors.

\begin{figure}[ht]
\centering
\begin{tabular}{c c}
\includegraphics[width=0.45\textwidth]{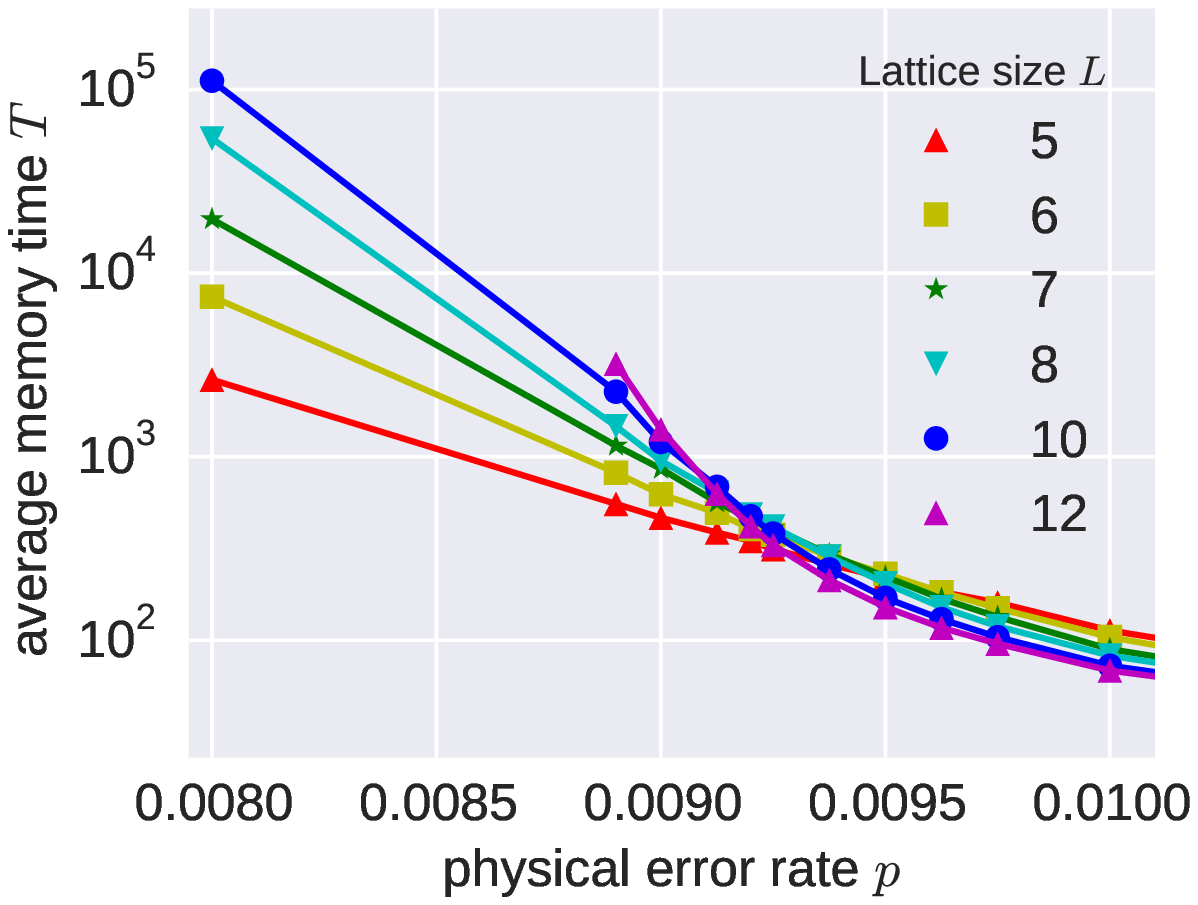} \hspace{0.25cm} & \includegraphics[width=0.45\textwidth]{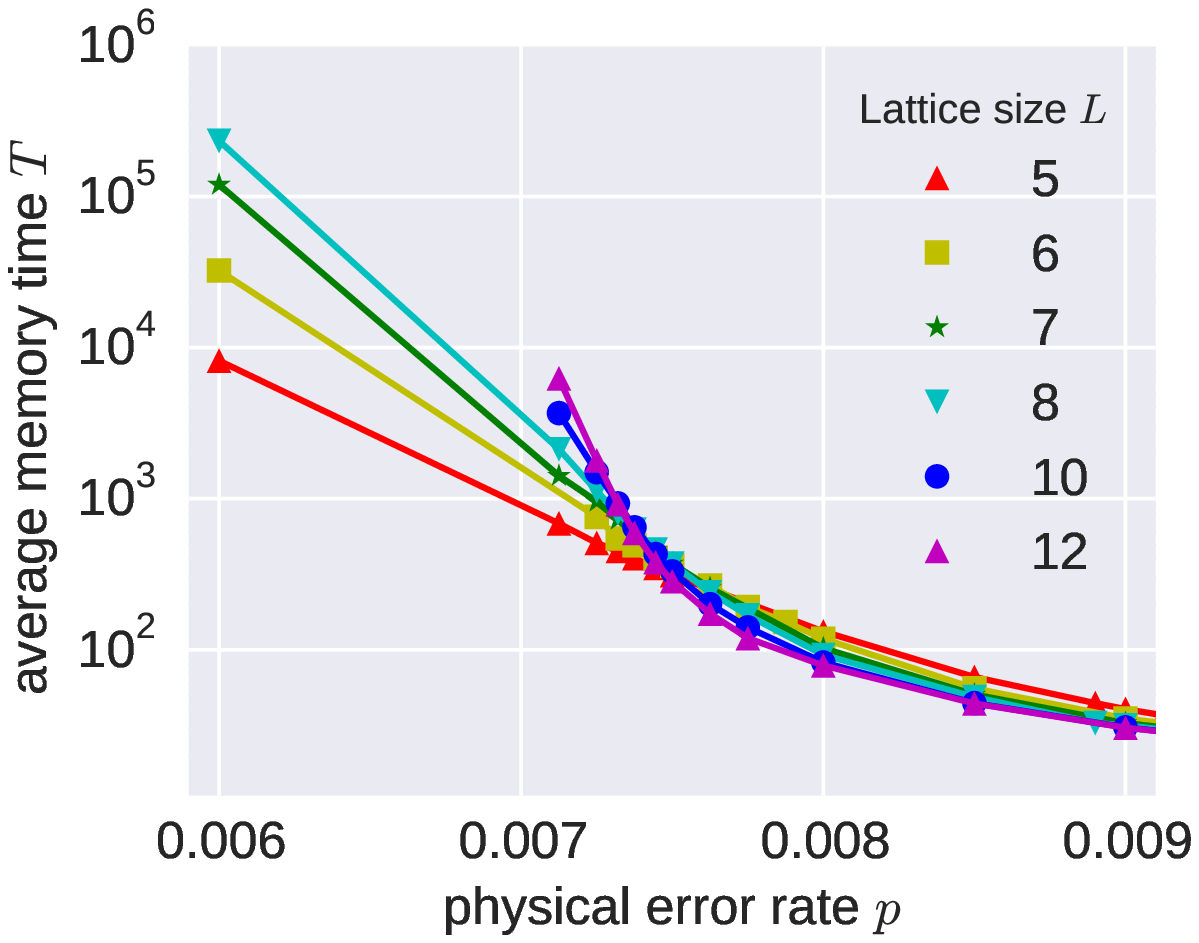}\\
(a)  &  (b)
\end{tabular}
\fcaption{(Color Online) Result for Toom's decoder.  At each time-step the update rule is applied to every plane once. (a) Without syndrome errors  (b) With syndrome errors }
\label{fig:toom_results} 
\end{figure}

In \cite{DKLP} the authors anticipated that a decoder based on Toom's rule will perform better than the DKLP decoder. Our results seem to confirm their intuition. In Fig.~\ref{fig:toom_results} we can see that for lattice sizes the cross-over points now lie in the region between $0.9\%$ and $1\%$ assuming perfect syndrome measurement and between $0.7\%$ and $0.8\%$ when including syndrome errors. Every data point in Fig. \ref{fig:toom_results} was obtained by $4\times 10^3$ trials.

For these cellular automaton decoders we have also observed that in the regime where the physical error rate is low the number of failures due to non-correctable syndromes is increasing with the size of the lattice (hence energy-barrier limited).

\begin{figure}[ht]
\centering
\includegraphics[width=0.5\textwidth]{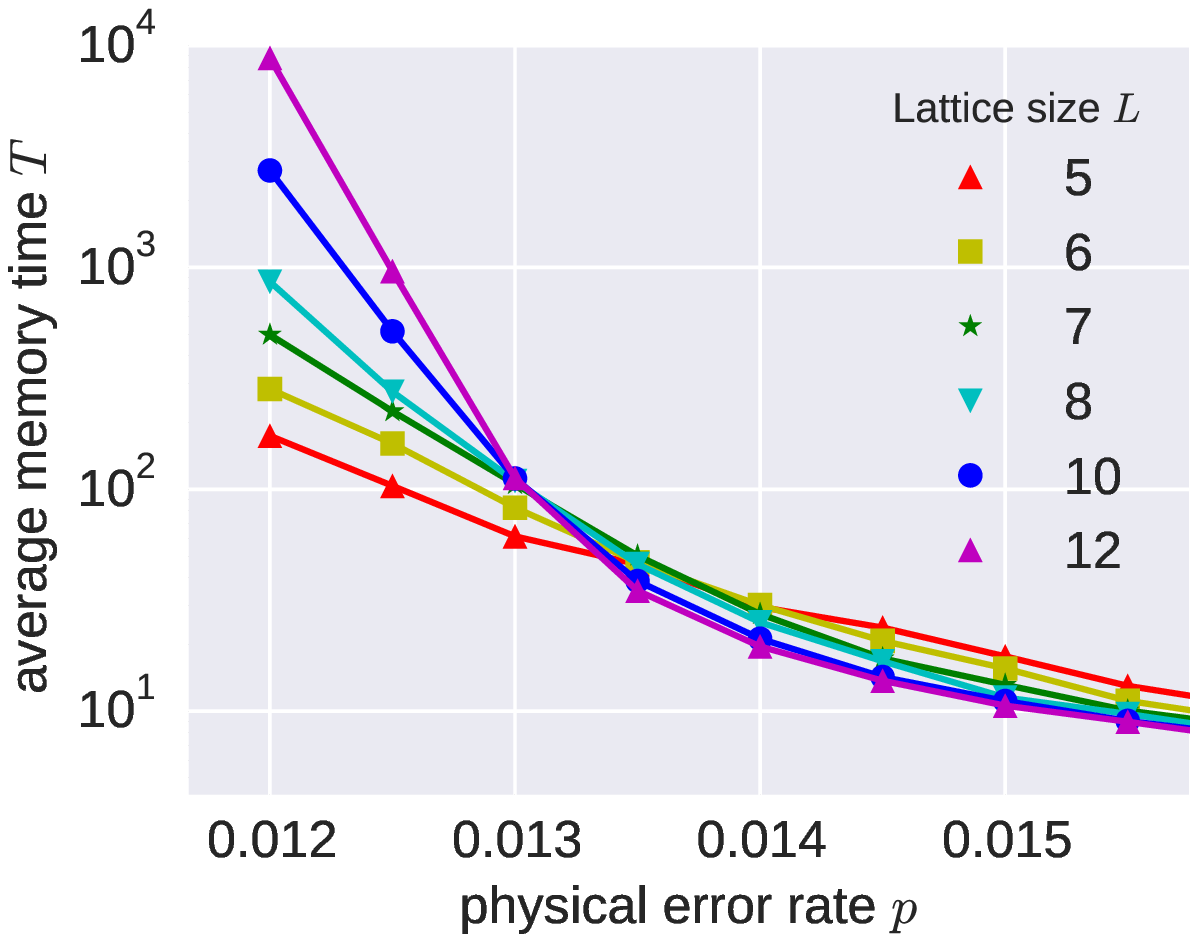}
\fcaption{Result for Toom's decoder with syndrome errors. At each time-step the update rule is applied to every plane 30 times.}
\label{fig:toom_results_repeated} 
\end{figure}

In the regime where classical computation is fast compared to the syndrome acquisition rate it is possible to apply the update rule multiple times. In the Hastings decoder we have similarly allowed to partition the lattice in neighborhoods and apply correction multiple times. In Fig.~\ref{fig:toom_results_repeated} we show the results when decoding using Toom's rule with noisy syndrome where we repeat the process of applying the update rule to every plane 30 times. We see that the performance of the decoder improves and that the crossover region is now between $1.25\%$ and $1.45\%$ which is close to the estimated value of the threshold for the Hastings decoder. Applying the update rule 100 times did not result in a further shift of the crossover region. Similarly, the memory time of the Toom's rule decoder is in the same regime as for the Hastings decoder (cf. Fig. \ref{fig:hastings_low_p}).

Overall, we thus find that the Toom's rule decoder performs quite well given the locality of the correction rules, in particular if one allows the rules to be applied multiple times in a QEC cycle. The Hastings decoder is more computationally intensive and applies more non-local error correction but its benefits as compared to Toom's rule are not clear given the current data.

The source code written for the numerical simulations can be downloaded from GitHub at \url{https://github.com/nikobreu/4Ddecoders} .

\section{Discussion}
\label{sec:discussion}

The Harrington decoder has poor thresholds as compared to RG, MWM or other non-local decoders. An open question is whether it can be improved by keeping its cellular automaton character at the lowest level but allowing for more computational power and faster communication. 
One idea would be to speed up the clock-rate of the logical higher level cells only, allowing for instantaneous correction and communication of syndrome information (or implementing MWM on level-1 syndromes). By construction of the decoder, the confidence of defects being the result of physical errors, instead of resulting from syndrome measurements errors, is substantially larger for higher level cells. We have observed that speeding up the decoder in the case of no measurement error has a very positive effect, hence assuming that the decoder is not limited by its functionality at the lowest level, one may expect some benefits. Alternatively one could further optimize the parameters $U$, $Q$, $f_n$ and $f_c$ by making them hierarchy-level dependent.
For the decoder to be practical, it should be altered to work on a planar surface code with less restrictive values for $L$.

The local single-shot decoders for the 4D toric code have noise thresholds which are still quite a bit below the $\approx 11\%$ conjectured optimal value.
Ways of handling narrow strips of errors or error syndromes which are locally of minimal length is to add randomness or temperature to the decoder rules.
We have attempted such a simulated annealing type procedure for the Hastings decoder (so that the dominant failure mode is no longer the presence of long error strings), but this did not improve the performance of the decoder.

\nonumsection{Acknowledgements}
\noindent
We thank Jim Harrington for initially providing us with his software implementation of the decoder. This work is supported by the European Research Council (EQEC, ERC Consolidator Grant No: 682726). We would like to thank Michael Kastyorano for valuable feedback on the 4D toric code results and Fernando Pastawski for discussing his results on the Toom's rule decoder.

\addcontentsline{toc}{section}{\protect\numberline{}Acknowledgements}

\section*{Appendix A: Local Update Rules}
\label{appendix: rules}
\addcontentsline{toc}{section}{\protect\numberline{}Appendix A}

Here we discuss in detail the local update rules used by the cellular automaton. These rules differ slightly from the rules described in the Harrington's thesis \cite{thesis:harrington}. As described in the main text the local update rules consists of 3 steps: 1) check nearest-neighbors, 2) check next-nearest-neighbors and 3) move to center.

The first step is summarized in Fig. \ref{im:subdivision}(a). A 0-cell will only perform a $Z$-flip on a qubit indicated by the red arrows. This prevents double flipping of the same qubit by two 0-cells. The direction of these arrows clearly depends on the location of the 0-cell in a colony. To make this precise, the colony is divided up into 8 parts: north-west quadrant (NWQ), north corridor (NC), north-east quadrant (NEQ), west corridor (WC), east corridor (EC), south-west quadrant (SWQ), south corridor (SC) and south-east quadrant (SEQ). As indicated in Fig. \ref{im:subdivision}(a), 0-cells in the corridors will only perform $Z$-flips on a single qubit whereas 0-cells in the quadrants can perform $Z$-flips on either one of two qubits. The western- and southern-most 0-cells are exceptional, they can perform $Z$-flips on more than 2 qubits, including a qubit {\em on the border} with another colony. If more than two nearest-neighbor 0-cells have a defect the following preference ordering is used. Priority is given to prevent error strings stretching across multiple colonies. Therefore the western and southern 0-cells will prefer to annihilate defects with their western and southern neighbors, respectively. Furthermore, priority is given to those neighbors which one can not flip to. For example, if a 0-cell in the NEQ detects defects at its southern and northern nearest-neighbor it will choose to annihilate its defect with its northern nearest-neighbor by not performing any $Z$-flip, since the qubit will already be flipped by its northern neighbor (unless it, on its turn, detects multiple defects at its directly neighboring 0-cells). Any other ambiguity is settled by the NESW rule: north before east before south before west.

\begin{figure}[htb]
\centering
\begin{tabular}{c c}
	\begin{tikzpicture}
	\draw [->, red, ultra thick] (1	,1) --  +(-0.5,0);
	\draw [->, red, ultra thick] (1	,2) --  +(-0.5,0);
	\draw [->, red, ultra thick] (1	,3) --  +(-0.5,0);
	\draw [->, red, ultra thick] (1	,4) --  +(-0.5,0);
	\draw [->, red, ultra thick] (1	,5) --  +(-0.5,0);	
	\draw [->, red, ultra thick] (1,1) --  +(0,-.5);
	\draw [->, red, ultra thick] (2,1) --  +(0,-.5);
	\draw [->, red, ultra thick] (3,1) --  +(0,-.5);
	\draw [->, red, ultra thick] (4,1) --  +(0,-.5);
	\draw [->, red, ultra thick] (5,1) --  +(0,-.5);
	\draw [very thin, gray, step=1.0cm] (0.5,0.5) grid (5.5,5.5);
	\draw [dashed, blue] (1.5,3) ellipse (1 and 0.5);
	\draw [->, red, ultra thick] (1,3) --  +(.5,0);
	\draw [->, red, ultra thick] (2,3) --  +(.5,0);
	\draw (1.5,3) node {WC};
	\draw [dashed, blue] (4.5,3) ellipse (1 and 0.5);
	\draw [->, red, ultra thick] (5,3) --  +(-.5,0);
	\draw [->, red, ultra thick] (4,3) --  +(-.5,0);
	\draw (4.5,3) node {EC};
	\draw [dashed, blue] (1.5,1.5) ellipse (1 and 1);
	\draw [->, red, ultra thick] (1,1) --  +(.5,0);
	\draw [->, red, ultra thick] (1,2) --  +(.5,0);
	\draw [->, red, ultra thick] (2,1) --  +(.5,0);
	\draw [->, red, ultra thick] (2,2) --  +(.5,0);
	\draw [->, red, ultra thick] (1,1) --  +(0,.5);
	\draw [->, red, ultra thick] (1,2) --  +(0,.5);
	\draw [->, red, ultra thick] (2,1) --  +(0,.5);
	\draw [->, red, ultra thick] (2,2) --  +(0,.5);
	\draw (1.5,1.5) node {SWQ};
	\draw [dashed, blue] (1.5,4.5) ellipse (1 and 1);
	\draw [->, red, ultra thick] (1,4) --  +(.5,0);
	\draw [->, red, ultra thick] (1,5) --  +(.5,0);
	\draw [->, red, ultra thick] (2,4) --  +(.5,0);
	\draw [->, red, ultra thick] (2,5) --  +(.5,0);
	\draw [->, red, ultra thick] (1,4) --  +(0,-.5);
	\draw [->, red, ultra thick] (1,5) --  +(0,-.5);
	\draw [->, red, ultra thick] (2,4) --  +(0,-.5);
	\draw [->, red, ultra thick] (2,5) --  +(0,-.5);
	\draw (1.5,4.5) node {NWQ};
	\draw [dashed, blue] (4.5,1.5) ellipse (1 and 1);
	\draw [->, red, ultra thick] (4,1) --  +(-.5,0);
	\draw [->, red, ultra thick] (4,2) --  +(-.5,0);
	\draw [->, red, ultra thick] (5,1) --  +(-.5,0);
	\draw [->, red, ultra thick] (5,2) --  +(-.5,0);
	\draw [->, red, ultra thick] (4,1) --  +(0,.5);
	\draw [->, red, ultra thick] (4,2) --  +(0,.5);
	\draw [->, red, ultra thick] (5,1) --  +(0,.5);
	\draw [->, red, ultra thick] (5,2) --  +(0,.5);
	\draw (4.5,1.5) node {SEQ};
	\draw [dashed, blue] (4.5,4.5) ellipse (1 and 1);
	\draw [->, red, ultra thick] (4,4) --  +(-.5,0);
	\draw [->, red, ultra thick] (4,5) --  +(-.5,0);
	\draw [->, red, ultra thick] (5,4) --  +(-.5,0);
	\draw [->, red, ultra thick] (5,5) --  +(-.5,0);
	\draw [->, red, ultra thick] (4,4) --  +(0,-.5);
	\draw [->, red, ultra thick] (4,5) --  +(0,-.5);
	\draw [->, red, ultra thick] (5,4) --  +(0,-.5);
	\draw [->, red, ultra thick] (5,5) --  +(0,-.5);
	\draw (4.5,4.5) node {NEQ};
	\draw [dashed, blue] (3,1.5) ellipse (0.5 and 1);
	\draw [->, red, ultra thick] (3,1) --  +(0,.5);
	\draw [->, red, ultra thick] (3,2) --  +(0,.5);
	\draw (3,1.5) node {SC};
	\draw [dashed, blue] (3,4.5) ellipse (0.5 and 1);
	\draw [->, red, ultra thick] (3,4) --  +(0,-.5);
	\draw [->, red, ultra thick] (3,5) --  +(0,-.5);
	\draw (3,4.5) node {NC};
	\end{tikzpicture} \hspace{0.5cm} & 
	\hspace{0.5cm}
		\begin{tikzpicture}
	\draw [->, red, ultra thick] (1,1) --  +(.4,.4);
	\draw [->, red, ultra thick] (1,1) --  +(.4,-.4);
	\draw [->, red, ultra thick] (1,1) --  +(-.4,-.4);
	\draw [->, red, ultra thick] (1,1) --  +(-.4,.4);
	\draw [->, red, ultra thick] (2,1) --  +(.4,.4);
	\draw [->, red, ultra thick] (2,1) --  +(.4,-.4);
	\draw [->, red, ultra thick] (2,1) --  +(-.4,-.4);
	\draw [->, red, ultra thick] (2,1) --  +(-.4,.4);
	\draw [->, red, ultra thick] (3,1) --  +(.4,.4);
	\draw [->, red, ultra thick] (3,1) --  +(.4,-.4);
	\draw [->, red, ultra thick] (3,1) --  +(-.4,-.4);
	\draw [->, red, ultra thick] (3,1) --  +(-.4,.4);
	\draw [->, red, ultra thick] (4,1) --  +(.4,.4);
	\draw [->, red, ultra thick] (4,1) --  +(.4,-.4);
	\draw [->, red, ultra thick] (4,1) --  +(-.4,-.4);
	\draw [->, red, ultra thick] (4,1) --  +(-.4,.4);
	\draw [->, red, ultra thick] (5,1) --  +(.4,.4);
	\draw [->, red, ultra thick] (5,1) --  +(.4,-.4);
	\draw [->, red, ultra thick] (5,1) --  +(-.4,-.4);
	\draw [->, red, ultra thick] (5,1) --  +(-.4,.4);
	\draw [->, red, ultra thick] (1,2) --  +(.4,.4);
	\draw [->, red, ultra thick] (1,2) --  +(.4,-.4);
	\draw [->, red, ultra thick] (1,2) --  +(-.4,-.4);
	\draw [->, red, ultra thick] (1,2) --  +(-.4,.4);
	\draw [->, red, ultra thick] (2,2) --  +(.4,.4);
	\draw [->, red, ultra thick] (2,2) --  +(.4,-.4);
	\draw [->, red, ultra thick] (2,2) --  +(-.4,.4);
	\draw [->, red, ultra thick] (3,2) --  +(.4,.4);
	\draw [->, red, ultra thick] (3,2) --  +(-.4,.4);
	\draw [->, red, ultra thick] (4,2) --  +(.4,.4);
	\draw [->, red, ultra thick] (4,2) --  +(-.4,-.4);
	\draw [->, red, ultra thick] (4,2) --  +(-.4,.4);
	\draw [->, red, ultra thick] (5,2) --  +(.4,.4);
	\draw [->, red, ultra thick] (5,2) --  +(.4,-.4);
	\draw [->, red, ultra thick] (5,2) --  +(-.4,-.4);
	\draw [->, red, ultra thick] (5,2) --  +(-.4,.4);
	\draw [->, red, ultra thick] (1,3) --  +(.4,.4);
	\draw [->, red, ultra thick] (1,3) --  +(.4,-.4);
	\draw [->, red, ultra thick] (1,3) --  +(-.4,-.4);
	\draw [->, red, ultra thick] (1,3) --  +(-.4,.4);
	\draw [->, red, ultra thick] (2,3) --  +(.4,.4);
	\draw [->, red, ultra thick] (2,3) --  +(.4,-.4);
	\draw [->, red, ultra thick] (4,3) --  +(-.4,-.4);
	\draw [->, red, ultra thick] (4,3) --  +(-.4,.4);
	\draw [->, red, ultra thick] (5,3) --  +(.4,.4);
	\draw [->, red, ultra thick] (5,3) --  +(.4,-.4);
	\draw [->, red, ultra thick] (5,3) --  +(-.4,-.4);
	\draw [->, red, ultra thick] (5,3) --  +(-.4,.4);
	\draw [->, red, ultra thick] (1,4) --  +(.4,.4);
	\draw [->, red, ultra thick] (1,4) --  +(.4,-.4);
	\draw [->, red, ultra thick] (1,4) --  +(-.4,-.4);
	\draw [->, red, ultra thick] (1,4) --  +(-.4,.4);
	\draw [->, red, ultra thick] (2,4) --  +(.4,.4);
	\draw [->, red, ultra thick] (2,4) --  +(-.4,-.4);
	\draw [->, red, ultra thick] (2,4) --  +(.4,-.4);
	\draw [->, red, ultra thick] (3,4) --  +(.4,-.4);
	\draw [->, red, ultra thick] (3,4) --  +(-.4,-.4);
	\draw [->, red, ultra thick] (4,4) --  +(.4,-.4);
	\draw [->, red, ultra thick] (4,4) --  +(-.4,-.4);
	\draw [->, red, ultra thick] (4,4) --  +(-.4,.4);
	\draw [->, red, ultra thick] (5,4) --  +(.4,.4);
	\draw [->, red, ultra thick] (5,4) --  +(.4,-.4);
	\draw [->, red, ultra thick] (5,4) --  +(-.4,-.4);
	\draw [->, red, ultra thick] (5,4) --  +(-.4,.4);
	\draw [->, red, ultra thick] (1,5) --  +(.4,.4);
	\draw [->, red, ultra thick] (1,5) --  +(.4,-.4);
	\draw [->, red, ultra thick] (1,5) --  +(-.4,-.4);
	\draw [->, red, ultra thick] (1,5) --  +(-.4,.4);
	\draw [->, red, ultra thick] (2,5) --  +(.4,.4);
	\draw [->, red, ultra thick] (2,5) --  +(.4,-.4);
	\draw [->, red, ultra thick] (2,5) --  +(-.4,-.4);
	\draw [->, red, ultra thick] (2,5) --  +(-.4,.4);
	\draw [->, red, ultra thick] (3,5) --  +(.4,.4);
	\draw [->, red, ultra thick] (3,5) --  +(.4,-.4);
	\draw [->, red, ultra thick] (3,5) --  +(-.4,-.4);
	\draw [->, red, ultra thick] (3,5) --  +(-.4,.4);
	\draw [->, red, ultra thick] (4,5) --  +(.4,.4);
	\draw [->, red, ultra thick] (4,5) --  +(.4,-.4);
	\draw [->, red, ultra thick] (4,5) --  +(-.4,-.4);
	\draw [->, red, ultra thick] (4,5) --  +(-.4,.4);
	\draw [->, red, ultra thick] (5,5) --  +(.4,.4);
	\draw [->, red, ultra thick] (5,5) --  +(.4,-.4);
	\draw [->, red, ultra thick] (5,5) --  +(-.4,-.4);
	\draw [->, red, ultra thick] (5,5) --  +(-.4,.4);
	\draw [red, ultra thick] (.8,1) arc (180:225:.2);
	\draw [red, ultra thick] (.8,2) arc (180:225:.2);
	\draw [red, ultra thick] (.8,3) arc (180:225:.2);
	\draw [red, ultra thick] (.8,3) arc (180:135:.2);
	\draw [red, ultra thick] (.8,4) arc (180:135:.2);
	\draw [red, ultra thick] (.8,5) arc (180:135:.2);
	\draw [red, ultra thick] (2,.8) arc (270:225:.2);
	\draw [red, ultra thick] (3,.8) arc (270:225:.2);
	\draw [red, ultra thick] (3,.8) arc (270:315:.2);
	\draw [red, ultra thick] (4,.8) arc (270:315:.2);
	\draw [red, ultra thick] (5,.8) arc (270:315:.2);
	\draw [red, ultra thick] (2,5.2) arc (90:135:.2);
	\draw [red, ultra thick] (3,5.2) arc (90:135:.2);
	\draw [red, ultra thick] (3,5.2) arc (90:45:.2);
	\draw [red, ultra thick] (4,5.2) arc (90:45:.2);
	\draw [red, ultra thick] (5,5.2) arc (90:45:.2);
	\draw [red, ultra thick] (5.2,2) arc (360:315:.2);
	\draw [red, ultra thick] (5.2,3) arc (360:315:.2);
	\draw [red, ultra thick] (5.2,3) arc (0:45:.2);
	\draw [red, ultra thick] (5.2,4) arc (0:45:.2);
	\draw [red, ultra thick] (1.2,1) arc (360:315:.2);
	\draw [red, ultra thick] (2.2,1) arc (360:315:.2);
	\draw [red, ultra thick] (1.2,2) arc (360:315:.2);
	\draw [red, ultra thick] (2.2,2) arc (360:315:.2);
	\draw [red, ultra thick] (1,1.2) arc (90:135:.2);
	\draw [red, ultra thick] (2,1.2) arc (90:135:.2);
	\draw [red, ultra thick] (1,2.2) arc (90:135:.2);
	\draw [red, ultra thick] (2,2.2) arc (90:135:.2);
	\draw [red, ultra thick] (1.2,4) arc (0:45:.2);
	\draw [red, ultra thick] (2.2,4) arc (0:45:.2);
	\draw [red, ultra thick] (1.2,5) arc (0:45:.2);
	\draw [red, ultra thick] (2.2,5) arc (0:45:.2);
	\draw [red, ultra thick] (1,3.8) arc (270:225:.2);
	\draw [red, ultra thick] (2,3.8) arc (270:225:.2);
	\draw [red, ultra thick] (1,4.8) arc (270:225:.2);
	\draw [red, ultra thick] (2,4.8) arc (270:225:.2);
	\draw [red, ultra thick] (3.8,4) arc (180:135:.2);
	\draw [red, ultra thick] (4.8,4) arc (180:135:.2);
	\draw [red, ultra thick] (3.8,5) arc (180:135:.2);
	\draw [red, ultra thick] (4.8,5) arc (180:135:.2);
	\draw [red, ultra thick] (4,3.8) arc (270:315:.2);
	\draw [red, ultra thick] (5,3.8) arc (270:315:.2);
	\draw [red, ultra thick] (4,4.8) arc (270:315:.2);
	\draw [red, ultra thick] (5,4.8) arc (270:315:.2);
	\draw [red, ultra thick] (3.8,1) arc (180:225:.2);
	\draw [red, ultra thick] (4.8,1) arc (180:225:.2);
	\draw [red, ultra thick] (3.8,2) arc (180:225:.2);
	\draw [red, ultra thick] (4.8,2) arc (180:225:.2);
	\draw [red, ultra thick] (5,1.2) arc (90:45:.2);
	\draw [red, ultra thick] (4,1.2) arc (90:45:.2);
	\draw [red, ultra thick] (5,2.2) arc (90:45:.2);
	\draw [red, ultra thick] (4,2.2) arc (90:45:.2);
	\draw [red, ultra thick] (1,1.2) arc (90:45:.2);
	\draw [red, ultra thick] (2,1.2) arc (90:45:.2);
	\draw [red, ultra thick] (3,1.2) arc (90:45:.2);
	\draw [red, ultra thick] (2,2.2) arc (90:45:.2);
	\draw [red, ultra thick] (3,2.2) arc (90:45:.2);
	\draw [red, ultra thick] (3,1.2) arc (90:135:.2);
	\draw [red, ultra thick] (4,1.2) arc (90:135:.2);
	\draw [red, ultra thick] (5,1.2) arc (90:135:.2);
	\draw [red, ultra thick] (3,2.2) arc (90:135:.2);
	\draw [red, ultra thick] (4,2.2) arc (90:135:.2);
	\draw [red, ultra thick] (1,4.8) arc (270:315:.2);
	\draw [red, ultra thick] (2,4.8) arc (270:315:.2);
	\draw [red, ultra thick] (3,4.8) arc (270:315:.2);
	\draw [red, ultra thick] (2,3.8) arc (270:315:.2);
	\draw [red, ultra thick] (3,3.8) arc (270:315:.2);
	\draw [red, ultra thick] (3,4.8) arc (270:225:.2);
	\draw [red, ultra thick] (4,4.8) arc (270:225:.2);
	\draw [red, ultra thick] (5,4.8) arc (270:225:.2);
	\draw [red, ultra thick] (3,3.8) arc (270:225:.2);
	\draw [red, ultra thick] (4,3.8) arc (270:225:.2);
	\draw [red, ultra thick] (1.2,2) arc (0:45:.2);
	\draw [red, ultra thick] (1.2,3) arc (0:45:.2);
	\draw [red, ultra thick] (2.2,3) arc (0:45:.2);
	\draw [red, ultra thick] (1.2,4) arc (360:315:.2);
	\draw [red, ultra thick] (1.2,3) arc (360:315:.2);
	\draw [red, ultra thick] (2.2,3) arc (360:315:.2);
	\draw [red, ultra thick] (4.8,2) arc (180:135:.2);
	\draw [red, ultra thick] (4.8,3) arc (180:135:.2);
	\draw [red, ultra thick] (3.8,3) arc (180:135:.2);
	\draw [red, ultra thick] (4.8,4) arc (180:225:.2);
	\draw [red, ultra thick] (4.8,3) arc (180:225:.2);
	\draw [red, ultra thick] (3.8,3) arc (180:225:.2);
	\draw [very thin, gray, step=1.0cm] (0.5,0.5) grid (5.5,5.5);
	\draw [dashed, blue] (1.5,3) ellipse (1 and 0.5);
	\draw (1.5,3) node {WC};
	\draw [dashed, blue] (4.5,3) ellipse (1 and 0.5);
	\draw (4.5,3) node {EC};
	\draw [dashed, blue] (1.5,1.5) ellipse (1 and 1);
	\draw (1.5,1.5) node {SWQ};
	\draw [dashed, blue] (1.5,4.5) ellipse (1 and 1);
	\draw (1.5,4.5) node {NWQ};
	\draw [dashed, blue] (4.5,1.5) ellipse (1 and 1);
	\draw (4.5,1.5) node {SEQ};
	\draw [dashed, blue] (4.5,4.5) ellipse (1 and 1);
	\draw (4.5,4.5) node {NEQ};
	\draw [dashed, blue] (3,1.5) ellipse (0.5 and 1);
	\draw (3,1.5) node {SC};
	\draw [dashed, blue] (3,4.5) ellipse (0.5 and 1);
	\draw (3,4.5) node {NC};
	\end{tikzpicture}\\
	(a)  &  (b)
\end{tabular}
\fcaption{(Color Online) Graphical representation of the local update rules. (a) Movement of defects if nearest-neighbor defects are present. Defects are moved depending on the relative location within a colony. The blue dotted ellipses indicate the subdivision of a $5\times 5$ colony into 4 quadrants (each containing 4 cells) and 4 corridors (each containing 2 cells). The red arrows indicate possible flipping directions. (Arrow tails indicate the cell flipping the qubit at the location of the arrow head.) (b) Movement of defects if next-nearest-neighbor defects are present. Red arrows indicate flipping directions by flipping the qubit on the link attached to the arrow by the red arc.}
\label{im:subdivision}
\end{figure}

The second step is summarized in Fig. \ref{im:subdivision}(b) and is very similar to the previous step except that the next-nearest-neighbors now play the role of the nearest-neighbors. Again, depending on their relative location, the 0-cells will only perform $Z$-flips on either 1 qubit (corridors) or 2 qubits (quadrants), with the exception now being all outer 0-cells. If a 0-cell detects a qubit at (a single) one of its next-nearest-neighbor 0-cells, indicated by the red arrows in Fig. \ref{im:subdivision}(b) it will perform a $Z$-flip on the qubit on the edge indicated by the arc attached to this arrow. In this way, for a pair of next-nearest-neighbor defects, if both 0-cells perform a $Z$-flip on a qubit, the action is coordinated in such a way that the pair of defects is annihilated in a single time step. Nearest-neighbor pairs for example oriented in SW-NE direction in the SWQ or NEQ are not annihilated in one time step since this will result in defects moving away from the colony center. Just as in the first step, first priority is given to annihilate with defects located in other colonies and second priority is to annihilate with defects one can not move to. Any other ambiguity is settled by the NESW rule.

The last step, moving towards the center is straightforward. For the corridors it means moving along the corridor. For the quadrants it means moving towards the furthest corridor, preferring the northern and southern direction if both corridors are equally far away.

Note that these rules differ slightly by those described in Harrington's thesis \cite{thesis:harrington}. Harrington gave priority to moving defects across colony borders, independent of detecting a defect at either a nearest-neighbor or a next-nearest-neighbor 0-cell. For example, if a southern 0-cell detects two defects, one at its SE-neighbor 0-cell and one at its N-neighbor 0-cell, our rules say to perform a $Z$-flip on its northern qubit, whereas the rules stated by Harrington say to perform a $Z$ flip on its southern qubit. We observed a slight improvement of average memory times with our rule set.
The rules can also be seen in action by running the decoder at the GitHub site.

\section*{References}
\noindent
\addcontentsline{toc}{section}{\protect\numberline{}References}


\begin{thebibliography}{10}

\bibitem{thesis:harrington}
J.~Harrington.
\newblock {\em Analysis of quantum error-correcting codes: symplectic lattice
  codes and toric codes}.
\newblock PhD thesis, CalTech, 2004.
\newblock \url{http://thesis.library.caltech.edu/1747/}.

\bibitem{hastings}
M.~B. Hastings.
\newblock Decoding in hyperbolic spaces: Quantum {LDPC} codes with linear rate
  and efficient error correction.
\newblock {\em Quant. Inf. Comput.}, 14(13-14):1187--1202, October 2014.

\bibitem{DKLP}
E.~Dennis, A.~Kitaev, A.~Landahl, and J.~Preskill.
\newblock Topological quantum memory.
\newblock {\em Journal of Mathematical Physics}, 43(9):4452--4505, 2002.

\bibitem{fowler2009high}
A.~Fowler, A.~Stephens, and P.~Groszkowski.
\newblock High-threshold universal quantum computation on the surface code.
\newblock {\em Phys. Rev. A}, 80(5):052312, 2009.

\bibitem{FMMC:review}
A.~Fowler, M.~Mariantoni, J.M. Martinis, and A.~N. Cleland.
\newblock Surface codes: Towards practical large-scale quantum computation.
\newblock {\em Phys. Rev. A}, 86:032324, 2012.

\bibitem{terhal:RMP}
B.M. Terhal.
\newblock Quantum error correction for quantum memories.
\newblock {\em Rev. Mod. Phys.}, 87:307--346, 2015.

\bibitem{fowler:O(1)}
A.~G. {Fowler}.
\newblock {Minimum weight perfect matching of fault-tolerant topological
  quantum error correction in average $O(1)$ parallel time}.
\newblock {\em Quant. Inf. Comp}, 15:0145--0158, 2015.

\bibitem{DP:3Ddecoding}
G.~{Duclos-Cianci} and D.~{Poulin}.
\newblock {Fault-tolerant renormalization group decoder for Abelian topological
  codes}.
\newblock {\em Quant. Inf. Comp.}, 14(9/10):0721--0740, 2014.

\bibitem{GACS198615}
P.~Gacs.
\newblock Reliable computation with cellular automata.
\newblock {\em Journal of Computer and System Sciences}, 32(1):15 -- 78, 1986.

\bibitem{guth_lubotzky}
L.~Guth and A.~Lubotzky.
\newblock Quantum error correcting codes and 4-dimensional arithmetic
  hyperbolic manifolds.
\newblock {\em Journal of Mathematical Physics}, 55(8):082202, 2014.

\bibitem{WHP:threshold}
C.~{Wang}, J.~{Harrington}, and J.~{Preskill}.
\newblock {Confinement-Higgs transition in a disordered gauge theory and the
  accuracy threshold for quantum memory}.
\newblock {\em Annals of Physics}, 303:31--58, 2003.

\bibitem{DP:fast}
G.~{Duclos-Cianci} and D.~{Poulin}.
\newblock {Fast Decoders for Topological Quantum Codes}.
\newblock {\em Phys. Rev. Lett.}, 104(5):050504, 2010.

\bibitem{BH_RGdecoder}
S.~Bravyi and J.~Haah.
\newblock Quantum self-correction in the {3D} cubic code model.
\newblock {\em Phys. Rev. Lett.}, 111:200501, Nov 2013.

\bibitem{2009PhRvB..79x5122H}
A.~{Hamma}, C.~{Castelnovo}, and C.~{Chamon}.
\newblock {Toric-boson model: Toward a topological quantum memory at finite
  temperature}.
\newblock {\em Phys. Rev. B}, 79(24):245122, June 2009.

\bibitem{Herold2015}
M.~Herold {\em et al.}
\newblock Cellular-automaton decoders for topological quantum memories.
\newblock {\em Npj Quantum Information}, 1:15010, 2015.

\bibitem{Herold_other}
M.~Herold {\em et al.}
\newblock Fault tolerant dynamical decoders for topological quantum memories
\newblock {\em ArXiv.org: 1511.05579}, November 2015.

%
%
%
%

\bibitem{fujii+:dissipative}
K.~Fujii, M.~Negoro, N.~Imoto, and M.~Kitagawa.
\newblock Measurement-free topological protection using dissipative feedback.
\newblock {\em Phys. Rev. X}, 4:041039, 2014.

\bibitem{DP_nonAbelian}
G.~{Dauphinais} and D.~{Poulin}.
\newblock {Fault-Tolerant Quantum Error Correction for non-Abelian Anyons}.
\newblock {\em ArXiv.org: 1607.02159}, July 2016.

\bibitem{takeda}
K.Takeda and H.Nishimori.
\newblock Self-dual random-plaquette gauge model and the quantum toric code.
\newblock {\em Nuclear Physics B}, 686(3):377 -- 396, 2004.

\bibitem{arakawa2005self}
G.~Arakawa, I.~Ichinose, T.~Matsui, and K.~Takeda.
\newblock {Self-duality and Phase Structure of the 4D Random-Plaquette $Z_2$
  Gauge Model}.
\newblock {\em Nuclear Physics B}, 709(1):296--306, 2005.

\bibitem{book:schrijver}
A.~Schrijver.
\newblock {\em Combinatorial Optimization: Polyhedra and Efficiency}
\newblock Springer-Verlag (2003).

\bibitem{toom}
A.L. Toom.
\newblock Stable and attractive trajectories in multicomponent systems.
\newblock {\em Advances in Probability}, 6:377 -- 396, 1980.

\bibitem{grinstein2004}
G.~Grinstein.
\newblock Can complex structures be generically stable in a noisy world?
\newblock {\em IBM Journal of Research and Development}, 48(1):5--12, 2004.

\bibitem{PCC:local}
F.~{Pastawski}, L.~{Clemente}, and J.~I. {Cirac}.
\newblock {Quantum memories based on engineered dissipation}.
\newblock {\em Phys. Rev. A}, 83(1):012304, 2011.

\bibitem{PastawskiPhD}
F.~Pastawski.
\newblock {\em Quantum memory: design and applications}.
\newblock PhD thesis, LMU, Munich, Germany, 2012.

\bibitem{AGP:ft}
P.~Aliferis, D.~Gottesman, and J.~Preskill.
\newblock Quantum accuracy threshold for concatenated distance-3 codes.
\newblock {\em Quant. Inf. Comp.}, 6:97--165, 2006.

\bibitem{gottesman:1D}
D.~{Gottesman}.
\newblock {Fault-tolerant quantum computation with local gates}.
\newblock {\em Journal of Modern Optics}, 47:333--345, 2000.

\bibitem{SDT:local}
K.M. Svore, D.P. DiVincenzo, and B.M. Terhal.
\newblock Noise threshold for a fault-tolerant two-dimensional lattice
  architecture.
\newblock {\em Quant. Inf. Comp.}, 7:297--318, 2007.

\bibitem{BNB:singleshot}
B.~J. {Brown}, N.~H. {Nickerson}, and D.~E. {Browne}.
\newblock {Fault-Tolerant Error Correction with the Gauge Color Code}.
\newblock {\em Nature Communications}, 7(12302), 2016.

\bibitem{BT:mem}
S.~Bravyi and B.~M. Terhal.
\newblock A no-go theorem for a two-dimensional self-correcting quantum memory
  based on stabilizer codes.
\newblock {\em New Journal of Physics}, 11:043029, 2009.

\end{thebibliography}
\end{document}